%% file: main.tex
\documentclass[aps,nofootinbib,prd,superscriptaddress, tightenlines,
notitlepage, twocolumn, floatfix]{revtex4-2}
\usepackage{tabularx}
\usepackage{graphicx}
\usepackage{epsfig}
\usepackage{amssymb,bm,tensor}
\usepackage{textcomp}
\usepackage{xcolor}
\usepackage{amsmath}
\usepackage[varg]{txfonts}
\usepackage{enumerate}
\usepackage{mathtools}
\usepackage[normalem]{ulem}
\usepackage{bm} % bold math
\usepackage[OT1]{fontenc}
\usepackage{multirow}
\usepackage{makecell}
\usepackage{soul}
\usepackage{booktabs}
\usepackage{nameref}
\usepackage[colorlinks,linkcolor=blue,citecolor=blue,urlcolor=blue ]{hyperref}

\input{aas_macros}

\begin{document}

\title{Dependence of postmerger properties on the thermal heating efficiency in neutron star mergers%gamma thermal
}
\author{Ming-Zhe Han}
\email{ming-zhe.han@aei.mpg.de}
\affiliation{Max-Planck-Institut f\"ur Gravitationsphysik (Albert-Einstein-Institut), Am M\"uhlenberg 1, D-14476 Potsdam-Golm, Germany}
\affiliation{Key Laboratory of Dark Matter and Space Astronomy, Purple Mountain Observatory, Chinese Academy of Sciences, Nanjing, 210033, People's Republic of China}

\author{Yong Gao}
\affiliation{Max-Planck-Institut f\"ur Gravitationsphysik (Albert-Einstein-Institut), Am M\"uhlenberg 1, D-14476 Potsdam-Golm, Germany}

\author{Kenta Kiuchi}
\affiliation{Max-Planck-Institut f\"ur Gravitationsphysik (Albert-Einstein-Institut), Am M\"uhlenberg 1, D-14476 Potsdam-Golm, Germany}
\affiliation{Center for Gravitational Physics and Quantum Information, Yukawa Institute for Theoretical Physics, Kyoto University, Kyoto, 606-8502, Japan}

\author{Masaru Shibata}
\affiliation{Max-Planck-Institut f\"ur Gravitationsphysik (Albert-Einstein-Institut), Am M\"uhlenberg 1, D-14476 Potsdam-Golm, Germany}
\affiliation{Center for Gravitational Physics and Quantum Information, Yukawa Institute for Theoretical Physics, Kyoto University, Kyoto, 606-8502, Japan}

\date{\today}

\begin{abstract}

We systematically perform numerical-relativity simulations for equal-mass binary neutron star mergers for the models varying the thermal index $\Gamma_{\rm th}$ with three different equations of state (EOSs) of the neutron stars (NSs), which are consistent with current multimessenger observational data and state-of-the-art theoretical calculations, and two different binary total mass ($m_0=2.7\ \text{and}\  2.9~M_\odot$). By varying the value of $\Gamma_{\rm th}$ within the hybrid EOS framework, we investigate the thermal effects on the merger dynamics, gravitational waves (GWs), and the dynamical mass ejection process. We find that the choice of the constant $\Gamma_{\rm th}$ can change the outcome of the remnant for specific EOSs and $m_0$. We also show that the dynamical ejecta mass is affected by the $\Gamma_{\rm th}$ value in a different way for different EOSs: for a stiff EOS the ejecta mass is high when $\Gamma_{\rm th}$ is small, while for softer EOSs the largest ejecta is achieved when $\Gamma_{\rm th} = 1.3$--$1.4$. While the inspiral motion does not depend on the $\Gamma_{\rm th}$ value, the postmerger phase evolution is highly affected by that. We show that the dominant peak frequency $f_2$ of the postmerger GW spectrum monotonically decreases as the $\Gamma_{\rm th}$ increases. We find that the universal relations between NS macroscopic properties and postmerger GW frequencies are subject to non-negligible thermal uncertainties, which can obscure the universal relation between the tidal deformability and $f_2$.
\end{abstract}

\maketitle

\section{Introduction}

Neutron stars (NSs) serve as exceptional laboratories for exploring the composition and state of matter at supranuclear densities~\cite{Lattimer:2012nd,Ozel:2016oaf,Lattimer:2015nhk,Oertel:2016bki,Lattimer:2021emm}. Due to the limited understanding of nonperturbative quantum chromodynamics (QCD), numerous zero-temperature equation of state (EOS) models have been developed within various theoretical frameworks, incorporating a range of possible compositions—from purely nucleonic matter to those including exotic degrees of freedom such as hyperons and quarks~(see, e.g.,~\cite{Lattimer:2000nx,Ozel:2016oaf} for reviews). These models are continuously studied and constrained by both experimental nuclear physics data~\cite{Reed:2021nqk,Hu:2021trw} and astrophysical observations~\cite{Hebeler:2013nza,Steiner:2010fz}. 

Currently, high-precision timing observations of binary pulsars have established a lower bound on the maximum mass of NSs at approximately $2\,M_{\odot}$~\cite{Demorest:2010bx,Antoniadis:2013pzd,Fonseca:2021wxt}, with multimessager data we can further constrain it to be $2.25^{+0.08}_{-0.07}~M_\odot$ \citep{Fan:2023spm} for nonrotating NS and $2.76^{+0.11}_{-0.09}~M_\odot$ \citep{2025arXiv250421408T} for Keplerian rotating NS (68.3\% credibility) by using Bayesian inference.
Pulse-profile modeling of several x-ray pulsars observed by the NICER satellite has provided constraints on both the mass and radius of NSs~\cite{Miller:2019cac,Riley:2019yda,Vinciguerra:2023qxq,Salmi:2024aum,Choudhury:2024xbk}. Meanwhile, the first detection of gravitational waves (GWs) from a binary neutron star (BNS) inspiral, GW170817~\cite{LIGOScientific:2017vwq}, gives valuable constraints on the EOS, primarily through the constraints on the NS's tidal deformability in the late inspiral phase~\cite{Hinderer:2007mb,Flanagan:2007ix,Hotokezaka:2016bzh,LIGOScientific:2018cki,LIGOScientific:2018hze}.

Due to the high Fermi energy of dense matter ($k_\mathrm{B}T_{\rm F} \sim 60$\,MeV where $k_\mathrm{B}$ is the Boltzmann's constant) at nuclear saturation density, observations of pulsars and BNS inspirals primarily probe the zero-temperature EOS in \(\beta\)-equilibrium. On the other hand, during and after BNS mergers, shock heating is expected to raise the temperature of the dense matter to 50--100\,MeV/$k_\mathrm{B}$~\cite{Oechslin:2006uk, Sekiguchi:2011zd, Sekiguchi:2011mc}, comparable to or well exceeding the Fermi energy. At such high temperatures, thermal pressure can contribute a significant fraction to the total pressure, thereby affecting various aspects of the merger dynamics---ranging from the lifetime of the merger remnant~\cite{Shibata:2005ss, Shibata:2006nm, Bauswein:2010dn, Paschalidis:2012ff} and the onset of fluid instabilities~\cite{DePietri:2018tpx,Camelio:2020mdi,Gao:2025nfj}, to the features of the postmerger GW spectrum and related quasiuniversal relations~\cite{Shibata:2005ss,Bauswein:2010dn,Figura:2020fkj,Raithel:2019gws,Raithel:2021hye,Raithel:2023zml,Blacker:2023opp,Miravet-Tenes:2024vba,Fields:2023bhs,Blacker:2023afl,Raithel:2022orm,Bauswein:2018bma,Breschi:2021xrx,Bernuzzi:2014kca,Bernuzzi:2015rla} as well as the amount and velocity distribution of the dynamical ejecta~\cite{Bauswein:2010dn,Raithel:2021hye,Raithel:2023zml}. The postmerger of BNSs thus provides us an experimental field to get some insights into the EOS at finite temperature.

Compared to zero-temperature EOSs, a much smaller number of finite-temperature EOSs based on microscopic nuclear theory are available in the literature (see, e.g.,~\cite{Lattimer:1991ads,Shen:1998gq,Typel:2013rza,Hempel:2011mk,Schneider:2017tfi}). A self-consistently constructed finite-temperature EOS typically depends on three thermodynamic parameters: rest-mass density \(\rho\), temperature \(T\), and electron fraction \(Y_e\). The thermal heating efficiency can be characterized by the local effective ``thermal index'' 
\begin{equation}
\Gamma_{\mathrm{th}}\left(\rho, T, Y_e\right)=\frac{P_{\mathrm{th}}\left(\rho, T, Y_e\right)}{\rho \varepsilon_{\mathrm{th}}\left(\rho, T, Y_e\right)}+1\,,
\end{equation}
where \(P_{\mathrm{th}}\) and \(\epsilon_{\mathrm{th}}\) are the thermal pressure and the thermal part of specific internal energy, respectively. 

The thermal index \(\Gamma_{\rm th}\) is very likely to have a strong density dependence~\cite{Constantinou:2014hha,Constantinou:2015mna,Carbone:2019pkr,Huth:2020ozf}, directly because of the density-dependence of the particle effective mass. For most nucleonic models, the thermal index \(\Gamma_{\rm th}\) varies between approximately \(1.3\) and \(2.0\) depending on the density and temperature, with relatively weak dependence on electron fraction \(Y_e\)~\cite{Raduta:2021coc}. The inclusion of exotic particles can significantly alter both the value and density dependence of \(\Gamma_{\rm th}\); in particular, many such models predict that $\Gamma_{\rm th} \lesssim 1.3$ across broad density ranges, and it can drop below zero when exotic particles become abundant~\cite{Kochankovski:2023trc,Kochankovski:2022rid,Raduta:2022elz}.

While the thermal behavior remains highly uncertain and is highly entangled with the zero-temperature composition and microphysical interactions, an essential first step involves (i) incorporating finite-temperature dependencies into different EOS models, and (ii) systematically quantifying, through merger simulations, how thermal heating efficiency impacts postmerger properties. Only through such studies can we provide reliable interpretations of multimessenger observations and move toward placing meaningful constraints on the EOS. State-of-the-art BNS merger simulations incorporate thermal effects with full temperature EOSs and neutrino transport~\cite{Sekiguchi:2011zd, Sekiguchi:2011mc, Sekiguchi:2015dma, Sekiguchi:2016bjd, Foucart:2016rxm, Radice:2021jtw,Foucart:2022kon}. However, due to the scarcity of fully temperature-dependent EOSs that capture the full range of thermal behaviors, as well as their relatively high computational cost, a parametric finite-temperature EOS is often useful for systematically performing merger simulations. 

To this end, many BNS simulations~\cite{Hotokezaka:2013iia,Kiuchi:2017pte,Read:2009yp,Bauswein:2010dn,Baiotti:2008ra} have adopted the so-called ``hybrid'' EOS approach~\cite{Janka:1993ads}, in which a thermal component—modeled as an ideal fluid with a constant thermal index $\Gamma_{\rm th}$—is added to a one-parameter cold EOS to account for shock heating effects. Reference~\cite{Shibata:2005ss} showed that, by varying $\Gamma_\mathrm{th}$ from 1.3 to 2 with a soft EOS for the cold part, the postmerger remnant and postmerger GW peak frequency can be significantly changed. 
Subsequently, Ref.~\cite{Bauswein:2010dn} also found that, by varying $\Gamma_{\rm th}$ from 1.5 to 2 while fixing the cold EOSs, the postmerger GW peak frequency can differ by about $50$--$250\,\rm Hz$. Reference~\cite{Figura:2020fkj} further investigated how different choices of $\Gamma_{\rm th}$ affect the GW signal and hydrodynamics. They suggest that $\Gamma_{\rm th}\approx 1.7$ best approximates the dynamical behavior of matter computed using fully finite-temperature EOS. To capture the strong density-dependence of $\Gamma_{\rm th}$,
Refs.~\cite{Raithel:2019gws,Raithel:2021hye,Raithel:2023zml} propose a new formalism based on the Landau's Fermi liquid theory for a special class of EOS models, in which the thermal contributions can be written purely in terms of the particle's effective mass. 
This formalism can approximate the merger dynamics and postmerger GWs more accurately~\cite{Raithel:2021hye,Raithel:2023zml,Raithel:2022MN}. However, it is worth noting that other classes of models can exhibit very different effective mass behavior with respect to density~\cite{Raduta:2021coc,Raduta:2022elz} and should be considered in future studies.

In this paper, we systematically investigate the impacts of finite-temperature effects on postmerger properties of BNS mergers by varying the thermal index $\Gamma_{\rm th}$ for a wide range within the hybrid framework for cold EOSs with different stiffness. 
By doing so, we aim to quantify the sensitivity of key postmerger observables, such as GW peak frequencies, (approximate) remnant lifetimes, and dynamical ejecta mass. Moreover, we study the systematic uncertainties induced by thermal components for universal relations connecting premerger and postmerger observables. 

The paper is organized as follows. In Sec.~\ref{sec2} we summarize our methods for this study. Section~\ref{sec3} presents numerical results paying particular attention to the dependence of the (approximate) lifetime of the merger remnant, dynamical ejecta mass, and gravitational waveforms on $\Gamma_\mathrm{th}$. Section~\ref{sec4} is devoted to a summary. Throughout this paper we use the units of $c=1$, $G=1$, and $k_\mathrm{B}=1$ where $c$, $G$, and $k_\mathrm{B}$ denote the speed of light, gravitational constant, and Boltzmann's constant, respectively.

\section{Methods}\label{sec2}

\subsection{Numerical relativity methods}

We perform hydrodynamics simulations of BNS mergers by using the \texttt{SACRA-MPI} code~\citep{Kiuchi:2017pte}, which solves the Einstein field equation in a moving puncture version~\citep{Campanelli:2005dd,Baker:2005vv,Marronetti:2007wz} of the Baumgarte-Shapiro-Shibata-Nakamura (BSSN) formulation~\citep{Shibata:1995we,Baumgarte:1998te} and locally incorporates the Z4c constraint-propagation prescription~\citep{Hilditch:2012fp, Kyutoku:2014yba}. 
Pure hydrodynamics equations are solved by using the Harten-Lax-van Leer contact (HLLC) solver~\citep{Mignone:2005ft,White:2015omx,Kiuchi:2022ubj}. 
To accelerate the computations, a hybrid MPI and OpenMP parallelization strategy is employed~\citep{Kiuchi:2017pte}. 
We do not take both magnetohydrodynamics and neutrino effects into account, as our analysis focuses only on the system's dynamics up to approximately 25~ms after the onset of merger, during which these effects are not expected to play a significant role on the postmerger evolution.

We implement a Berger--Oliger-type adaptive mesh refinement (AMR) algorithm to extend the simulation domain into the local GW zone while maintaining high spatial resolution near the NSs \citep{Berger:1984zza}. 
The AMR domain consists of 10 nested Cartesian refinement levels (from level 0 to 9), where each finer level has half the grid spacing of its parent level. 
The resolution at refinement level $l$ is given by $\Delta_l = L / (2^l N)$, with $l = 0, 1, ..., 9$ and $L$ denoting the distance from the coordinate origin to the outer boundary along each axis. 
Each level contains $(2N + 2) \times (2N + 2) \times (N + 1)$ cell-centered staggered grid points and $N$ is an even integer. 
Levels 0--5 prepare a single patch centered at the coordinate origin, while levels 6--9 have two comoving patches that follow the inspiral motion of the two NSs.

We use the public code \texttt{FUKA}~\citep{Papenfort:2021hod} to generate quasi-equilibrium initial data for the equal-mass BNS systems. The $3.5$ post-Newtonian order approximation of the orbital frequency $\Omega$ and the infall velocity $\dot{a}$ are used to reduce the orbital eccentricity to the order $10^{-3}$.

\subsection{EOS}

In order to span a larger range of unknown physics and to accelerate the computation for systematic studies, we employ completely nonparametric EOSs in the hybrid EOS framework. Specifically, the specific internal energy and pressure are decomposed to 
\begin{eqnarray}
  \varepsilon &=& \varepsilon_{\rm cold} + \varepsilon_{\rm th}\,, \\
  P &=& P_{\rm cold} + P_{\rm th}\,,
\end{eqnarray}
where the subscript ``cold'' indicates that the quantity is calculated at zero-temperature while the ``th'' indicates the thermal part accounting for shock heating in the system. 

The cold EOS only depends on rest-mass density $\rho$ and is parametrized by a spectral expansionlike \citep{Lindblom:2013kra} feed forward neural network (FFNN) model in our work (see Refs.~\citep{Han:2021kjx,Han:2022rug} for more details). 
Following the method in Ref.~\citep{Han:2021kjx}, 
the squared sound speed is described by a function of the rest-mass density in the FFNN model as follows; 
\begin{eqnarray}
    y_{1,j} &=& \tanh(w_{0,j} \log\rho + b_{0,j})\,, \\
    y_{2,i} &=& \tanh\left(\sum^{N_1}_{j=1} w_{1,ij} y_{1,j} + b_{1,i}\right)\,, \\
    c_s^2 &=& S\left(\sum^{N_2}_{i=1} w_{2,i} y_{2,i} + b_2 \right)\,,
\end{eqnarray}
where $w_{0,j}$, $w_{1,ij}$, and $w_{2,i}$ are the weights while $b_{0,j}$, $b_{1,i}$, and $b_2$ are bias of the FFNN model, respectively. 
The hyper parameters $N_1$ and $N_2$ are both set to be 16 in this work. 
As for the activation functions, we use {\it hyperbolic tangent} and {\it sigmoid} functions, which are defined by
\begin{eqnarray}
    \tanh(x) &=& \frac{e^x - e^{-x}}{e^x + e^{-x}}\,,\\
    S(x) &=& \frac{1}{1 + e^{-x}}\,.
\end{eqnarray}

\begin{figure*}
    \centering
    \includegraphics[width=1.99\columnwidth]{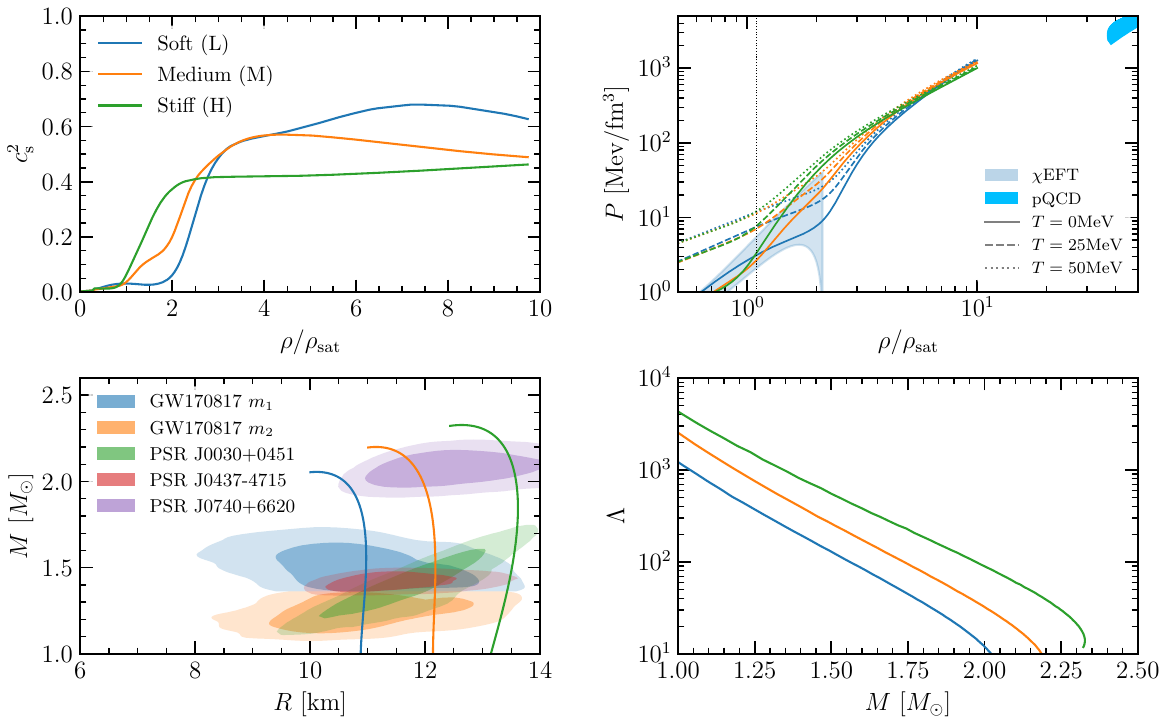}
    \caption{
    Three EOSs employed in this work: The upper panels show the squared sound speed $c_{\rm s}^2$ (left) and the total pressure $P$ (right) as functions of the rest-mass density $\rho$, normalized by the nuclear saturation density $\rho_{\rm sat}=2.7\times10^{14}{\rm g/cm^3}$, at different temperatures. In the upper right panel, the shaded regions indicate the pressure ranges predicted by $\chi$EFT and perturbative QCD, respectively. The vertical dashed line marks $\rho = 1.1~\rho_{\rm sat}$, the maximum density up to which $\chi$EFT constraints are applied. The different line styles stand for the different temperatures. The lower left panel displays the mass–radius relations, with the shaded bands representing the 68\% and 95\% credible regions from NS observational constraints. The lower right panel shows the dimensionless tidal deformability $\Lambda$ as a function of NS mass.}
    \label{fig:eos}
\end{figure*}

Using the model described above, we can randomly draw a set of weights and bias parameters from the standard normal distributions and then generate an EOS. 
The randomly generated EOSs are only theoretically realistic, so they may not be consistent with the various current constraints. 
Therefore, we then use the multimessenger data and state-of-the-art theoretical calculations to give further constraints on the EOSs.
In this work, we consider the following constraints:
\begin{itemize}
    \item Pressure constraints up to 1.1~$\rho_{\rm sat}$ from the next-to-next-to-next-to leading order chiral effective field theory ($\chi$EFT) calculations \citep{Drischler:2017wtt,Drischler:2020hwi}.
    \item Pressure constraints of the high-density side of the EOSs extrapolated from the perturbative QCD calculations \citep{Kurkela:2009gj,Gorda:2021znl} at very high ($\sim40~\rho_{\rm sat}$) density.
    \item Mass and tidal deformability measurements from GW170817 \citep{LIGOScientific:2017vwq}, and the mass-radius constraints \citep{LIGOScientific:2018cki} inferred from that.
    \item The simultaneous measurements of mass and radius from the NICER observations, i.e., the PSR 0030+0451 \citep{Vinciguerra:2023qxq}, PSR 0740+6620 \citep{Salmi:2024aum}, and PSR 0437-4715 \citep{Choudhury:2024xbk}.
\end{itemize}
After applying the above constraints, we can get an ensemble of the EOSs, and then, we select three of them which range from soft to stiff in this work.
Figure~\ref{fig:eos} shows the information for the selected EOSs with the constraints.
Three EOSs, i.e., soft, medium, and stiff, are selected based on their radii and tidal deformabilities of a canonical $1.4~M_\odot$ NS.
The properties of these EOSs are listed in Tab.~\ref{tab:EOSs}.

\begin{table}[htbp]
  \centering
  
  \caption{Properties of the EOSs used in this work. The maximum mass for non-rotating cold NSs, $M_{\rm TOV}$, the radius and dimensionless tidal deformability of a 1.4 $M_\odot$ NS, $M_{1.4}$ and $\Lambda_{1.4}$, and the cold pressure at twice nuclear saturation densities, $P_{\rm cold}(2\rho_{\rm sat})$.}
  \label{tab:EOSs}
  
  \begingroup
  \setlength{\tabcolsep}{5pt} % Default value: 6pt
  \renewcommand{\arraystretch}{1} % Default value: 1
  
  \begin{tabular}{l ccc c}
    \hline
    \hline
    EOS & $M_{\rm TOV}~[M_\odot]$ & $R_{1.4}~[{\rm km}]$ & $\Lambda_{1.4}$ & $P_{\rm cold}(2\rho_{\rm sat})~{\rm [MeV/fm^3]}$ \\
    \hline
    Soft   & $2.05$ & $10.96$ & $196$ & $7.67$ \\
    Medium & $2.20$ & $12.18$ & $397$ & $19.82$ \\
    Stiff  & $2.32$ & $13.45$ & $800$ & $40.80$ \\
    \hline
    \hline
  \end{tabular}
  \endgroup
\end{table}

In the hybrid EOS framework, the thermal part is treated by the ideal gas EOS with a constant thermal index $\Gamma_{\rm th}$, i.e.,
\begin{equation}
  \label{eq:p_th}
      P_{\rm th} = (\Gamma_{\rm th} - 1)\rho \varepsilon_{\rm th} \,.
  \end{equation}
With this prescription, the temperature can be calculated by
\begin{equation}
    T = m_{\rm n} (\Gamma_{\rm th} - 1) \varepsilon_{\rm th},
\end{equation}
where $m_{\rm n}$ = 931.19 MeV is the necleon mass.
The $\Gamma_{\rm th}$ value determines {\it the conversion efficiency of kinetic energy to thermal energy in shocks} and is the most important parameter in our work.
We perform many simulations by employing a broad range of $\Gamma_{\rm th}$, which will be described in detail in Sec.~\ref{sec:model}.

\subsection{Model}
\label{sec:model}

\begin{table}[htbp]
  \centering
  
  \caption{Parameters for the initial conditions adopted in our numerical simulations. The name of models and EOSs, the mass of the NS $M$, the radius of the NS $R$, the dimensionless tidal deformability of the NS $\Lambda$, the thermal index $\Gamma_{\rm th}$, and the grid spacings $\Delta_{x}$ for the highest-resolution domain with different grid resolutions (N102, N82, N62); the highest resolution model N102 is missing in some models.}
  \label{tab:models}
  
  \begingroup
  \setlength{\tabcolsep}{2.6pt} % Default value: 6pt
  \renewcommand{\arraystretch}{1} % Default value: 1
  
  \begin{tabular}{l ccc ccc}
    \hline
    \hline
    Model & EOS & $M~[M_{\odot}]$ & $R~[{\rm km}]$ & $\Lambda$ & $\Gamma_{\rm th}$ & $\Delta_{x}~{\rm [m]}$\\
    \hline
    M-M27-G11 & medium & 1.35  & 12.17 & 488 & 1.1 & $(147,182,240)$ \\
    M-M27-G12 & medium & 1.35  & 12.17 & 488 & 1.2 & $(182,240)$ \\
    M-M27-G13 & medium & 1.35  & 12.17 & 488 & 1.3 & $(147,182,240)$ \\
    M-M27-G14 & medium & 1.35  & 12.17 & 488 & 1.4 & $(182,240)$ \\
    M-M27-G15 & medium & 1.35  & 12.17 & 488 & 1.5 & $(147,182,240)$ \\
    M-M27-G16 & medium & 1.35  & 12.17 & 488 & 1.6 & $(182,240)$ \\
    M-M27-G17 & medium & 1.35  & 12.17 & 488 & 1.7 & $(147,182,240)$ \\
    M-M27-G18 & medium & 1.35  & 12.17 & 488 & 1.8 & $(182,240)$ \\
    M-M27-G19 & medium & 1.35  & 12.17 & 488 & 1.9 & $(147,182,240)$ \\
    M-M27-G20 & medium & 1.35  & 12.17 & 488 & 2.0 & $(182,240)$ \\
    M-M29-G11 & medium & 1.45  & 12.18 & 320 & 1.1 & $(147,182,240)$ \\
    M-M29-G12 & medium & 1.45  & 12.18 & 320 & 1.2 & $(182,240)$ \\
    M-M29-G13 & medium & 1.45  & 12.18 & 320 & 1.3 & $(147,182,240)$ \\
    M-M29-G15 & medium & 1.45  & 12.18 & 320 & 1.5 & $(147,182,240)$ \\
    M-M29-G16 & medium & 1.45  & 12.18 & 320 & 1.6 & $(182,240)$ \\
    M-M29-G17 & medium & 1.45  & 12.18 & 320 & 1.7 & $(147,182,240)$ \\
    M-M29-G18 & medium & 1.45  & 12.18 & 320 & 1.8 & $(182,240)$ \\
    M-M29-G19 & medium & 1.45  & 12.18 & 320 & 1.9 & $(147,182,240)$ \\
    M-M29-G20 & medium & 1.45  & 12.18 & 320 & 2.0 & $(182,240)$ \\
    \hline
    L-M27-G11 & soft & 1.35 & 10.95 & 242 & 1.1 & $(132,164,216)$ \\
    L-M27-G15 & soft & 1.35 & 10.95 & 242 & 1.5 & $(132,164,216)$ \\
    L-M27-G20 & soft & 1.35 & 10.95 & 242 & 2.0 & $(132,164,216)$ \\
    L-M29-G11 & soft & 1.45 & 10.97 & 158 & 1.1 & $(132,164,216)$ \\
    L-M29-G15 & soft & 1.45 & 10.97 & 158 & 1.5 & $(132,164,216)$ \\
    L-M29-G20 & soft & 1.45 & 10.97 & 158 & 2.0 & $(132,164,216)$ \\
    \hline
    H-M27-G11 & stiff & 1.35 & 13.42 & 974 & 1.1 & $(162,200,264)$ \\
    H-M27-G15 & stiff & 1.35 & 13.42 & 974 & 1.5 & $(162,200,264)$ \\
    H-M27-G20 & stiff & 1.35 & 13.42 & 974 & 2.0 & $(162,200,264)$ \\
    H-M29-G11 & stiff & 1.45 & 13.48 & 665 & 1.1 & $(162,201,265)$ \\
    H-M29-G15 & stiff & 1.45 & 13.48 & 665 & 1.5 & $(162,201,265)$ \\
    H-M29-G20 & stiff & 1.45 & 13.48 & 665 & 2.0 & $(162,201,265)$ \\
    \hline
    \hline
  \end{tabular}
  \endgroup
\end{table}

We perform simulations for $\Gamma_{\rm th}=1.1$--$2.0$, with three representative EOSs: soft, medium, and stiff, which are described in the previous subsection.
The value of $\Gamma_{\rm th}$ is densely chosen for the medium EOS, while for the soft and stiff EOSs, simulations are performed with $\Gamma_{\rm th} = 1.1$, 1.5, and 2.0. % to study the influence of the EOS in comparison to thermal effects. 
To investigate the different outcomes of merger remnants and their collapse behavior within the simulation timescale, we choose total binary masses of $m_0 = 2.7$ and $2.9~M_\odot$. 
Table~\ref{tab:models} lists the equal-mass BNS models and their initial parameters used in this work. 
Each model is labeled as EOS-$m_0$-$\Gamma_{\rm th}$; e.g., M-M27-G11 refers to a model with the ``medium'' EOS, $m_0 = 2.7~M_\odot$, and $\Gamma_{\rm th} = 1.1$. 
``L" and ``H" refer to the soft and stiff EOSs, respectively.

To further verify our numerical results, we set multiple grid resolutions: N62, N82, and N102 (not for all models) for each model, which correspond to approximately 50, 66, and 82 grid points across the NS radius on the finest AMR level, respectively. For example, for the medium EOS with $R_{1.35} = 12.17$~km, the corresponding finest AMR resolutions are 240~m, 180~m, and 150~m, from low to high resolutions.
Throughout the rest part of this paper, all the tables and figures are produced by the results with N82 resolution unless otherwise specified.

\subsection{Diagnostics}

\subsubsection{Gravitational wave extraction}

The gravitational waveforms \citep{Yamamoto:2008js} are extracted from the complex Weyl scalar $\Psi_4$, which can be written as the expansion
\begin{equation}
    \Psi_4\left(t_{\mathrm{ret}}, r_0, \theta, \phi\right)=\sum_{l, m} \Psi_4^{l, m}\left(t_{\mathrm{ret}}, r_0\right) _{-2}Y_{l m}(\theta, \phi),
\end{equation}
where $_{-2}Y_{l m}(\theta, \phi)$ is the spin-weighted spherical harmonics of weight $-2$ and $t_{\rm ret}$ is the retarded time.
We use Nakano's method \citep{Nakano:2015rda} to obtain $\Psi_4^{l, m}$ at infinity from $\Psi_4^{l,m}(t_\mathrm{ret},r_0)$ that is extracted at a fixed radius $r_0$ (in this work $r_0=480\,M_\odot$) as
\begin{align}
      D \Psi_4^{l,m,\infty}(t_\mathrm{ret}) & \equiv C(r_0)\Big[D \Psi_4^{l,m}(t_\mathrm{ret},r_0) \nonumber\\
    &  - \frac{(l-1)(l+2)}{2}\int^{t_\text{ret}} \Psi_4^{l,m}(t',r_0) dt' \Big]\,,
\end{align}
where $C(r_0)$ is a function of the extraction radius $r_0$.
Since our coordinate is similar to an isotropic coordinate of a nonrotating black hole (BH) in the wave zone, we set $D \approx r_0 [1 + m_0/(2r_0)]$ and $C(r_0)=1 - 2m_0/D$ \citep{Kiuchi:2017pte,Kiuchi:2019kzt}, and in turn define the retarded time as
\begin{equation}
    t_{\mathrm{ret}} \equiv t - \left[ D + 2m_0 \ln\left( \frac{D}{2m_0} - 1 \right) \right]\,.
\end{equation}

By integrating $\Psi_4^{l, m, \infty}$ twice in time, we can obtain the strain of GWs 
\begin{align}
    h^{l,m,\infty}(t_\mathrm{ret}) &= h^{l,m,\infty}_+ (t_\mathrm{ret}) - i h^{l,m,\infty}_\times(t_\mathrm{ret}) \nonumber\\
  &= - \int^{t_\mathrm{ret}} dt'\int^{t'} \Psi_4^{l,m,\infty}(t'')dt'' \,.
\end{align}
We employ the fixed frequency method \citep{Reisswig:2010di} for the time integration of the waveforms,
\begin{equation}
    h^{l,m,\infty}(t_\mathrm{ret}) = \int df' \frac{\tilde{\Psi}_4^{l,m,\infty}(f')}{(2\pi\max[f',f_\mathrm{cut}])^2} \exp(2\pi i f' t_\mathrm{ret})\,,
\end{equation}
where $\tilde{\Psi_4^{l, m, \infty}}(f)$ is the Fourier component of $\Psi_4^{l, m, \infty}(t)$ and $f_{\mathrm{cut}}$ is the cutoff frequency when performing Fourier transformation, which is chosen as $0.8m\Omega_0/(2\pi)$ in our work, where $\Omega_0$ is the initial orbital angular velocity and $m$ denotes the order of spherical harmonics. 

GWs for each $(l,m)$ mode can be written as the combination of amplitude and phase, i.e.,
\begin{equation}
    h^{l,m,\infty}(t_\mathrm{ret}) = A^{l,m,\infty}(t_\mathrm{ret}) e^{-i\Phi^{l,m}(t_\mathrm{ret})}\,.
\end{equation}
The instantaneous GW frequency defined by $d\Phi^{l, m}/dt_{\mathrm{ret}}$ can be calculated as
\begin{equation}
    f_\mathrm{GW} = \frac{1}{2\pi}{\rm Im} \left(\frac{h^{*2,2,\infty}\dot{h}^{2,2,\infty}}{|h^{2,2,\infty}|^2}\right)\,,
\end{equation}
where $h^{*2,2,\infty}$ is the complex conjugate of $h^{2,2,\infty}$.
To account for the GW power spectrum, we define the effective amplitude as:
\begin{equation}
    h_{\rm eff}(f) \equiv f\tilde{h}(f) = f\sqrt{\frac{|\tilde{h}_+^{2,2}(f)|^2 + |\tilde{h}_{\times}^{2,2}(f)|^2}{2}}\,,
\end{equation}
where $\tilde{h}_{+/\times}^{2,2}(f)$ is the Fourier transform of the $h_{+/\times}^{2,2}(t)$.

\subsubsection{Ejecta and disk}

After the merger of BNS, the matter can be classified into three categories: the matter that is bounded either in the dense core of a remnant NS or directly swallowed into a BH, the matter that remains bound and forms an accretion disk around the remnant, and the matter that escapes from the system as ejecta. To analyze the postmerger dynamics, we calculate the properties of the accretion disk and the ejecta, which are critical for understanding the electromagnetic emissions associated with BNS mergers.

The ejected matter is defined as the fluid elements that become unbound from the gravitational potential of the merger remnant.
To identify the unbound matter, we evaluate the quantity $u_\mu t^\mu = u_t$ at each grid point, where $t^\mu=(1,0,0,0)$ is a timelike Killing vector at spatial infinity.
The ejecta are then defined using the geodesic criterion: matter that satisfies $ -u_t > 1$ is considered unbound.

To compute the mass of the ejecta, we integrate the conserved rest-mass density over the region satisfying this unbound condition:
\begin{equation}
    M_{\mathrm{eje}} = \int_{-u_t>1,\,r>r_{\mathrm{AH}}} \rho_*\, d^3x\,,
\end{equation}
where $\rho_*$ denotes the conserved rest-mass density, and $r_{\mathrm{AH}} = r_{\mathrm{AH}}(\theta, \varphi)$ is the coordinate radius of the apparent horizon, expressed as a function of the polar angles $\theta$ and $\varphi$, in a BH-centered coordinate frame ($r_\mathrm{AH}=0$ for the case that the remnant is a NS).

To estimate the average velocity of the ejecta, we first define its energy (sum of the rest-mass, internal, and kinetic energies) as
\begin{equation}
    E_{\mathrm{eje}} = \int_{-u_t>1,\,r>r_{\mathrm{AH}}} \rho_* e_0\, d^3x\,,
\end{equation}
where $e_0 = hw-P/(\rho w)$ is the specific energy in general relativity, with $h$ and $w$ being the specific enthalpy and the Lorentz factor, respectively.

The internal energy of the ejecta is given by
\begin{equation}
    U_{\mathrm{eje}} = \int_{-u_t>1,\,r>r_{\mathrm{AH}}} \rho_* \varepsilon\, d^3x\,,
\end{equation}
where $\varepsilon$ is the specific internal energy.

With these quantities, the kinetic energy of the ejecta is computed as
\begin{equation}
    T_{\mathrm{eje}} = E_{\mathrm{eje}} - U_{\mathrm{eje}} - M_{\mathrm{eje}}\,.
\end{equation}
Finally, assuming Newtonian dynamics at a far zone, the average velocity of the ejecta is estimated by
\begin{equation}
    v_{\mathrm{eje}} = \sqrt{\frac{2 T_{\mathrm{eje}}}{M_{\mathrm{eje}}}}\,.
\end{equation}

\section{Results}\label{sec3}

\begin{table*}[htbp]
  \centering
  
  \caption{Summary of numerical results: The second and third columns show the type of the remnant and the (approximate) lifetime of the  massive NS (MNS) temporarily formed, $\tau_{H}$, respectively. The following columns show the maximum rest-mass density $\rho_{\rm max}$, the ejecta mass $M_{\rm eje}$, the average velocity of ejected material $\bar{v}_{\rm eje}$. The last two columns show the frequency at the merger time $f_{\rm peak}$ and the dominant frequency of the postmerger GW spectrum $f_2$. The errors are estimated by the different resolutions, i.e., for a specific model (two or three different resolutions); we consider the result of the N82 model as the median value and take the difference between it and the results of other resolution(s). The (maximum) absolute value of the difference(s) is considered to be the symmetric error.}
  \label{tab:results}
  
  \begingroup
  \setlength{\tabcolsep}{14pt} % Default value: 6pt
  \renewcommand{\arraystretch}{1} % Default value: 1
  
  \begin{tabular}{l ccc ccc cc}
    \hline
    \hline
    Model & Type & $\tau_{H}$ [ms] & $\rho_{\rm max}/\rho_{\rm sat}$ & $M_{\rm eje}~[10^{-3}M_{\odot}]$ & $\bar{v}_{\rm eje}$ & $f_{\rm peak}~{\rm[kHz]}$ & $f_2~{\rm [kHz]}$ \\
    \hline
    M-M27-G11 & III &$> 25$ & $4.0$ & $3\pm1$ & $0.23$ & $1.86\pm0.04$ & $3.17\pm0.03$ \\
    M-M27-G12 & III &$> 25$ & $4.0$ & $13\pm1$ & $0.23$ & $1.87\pm0.11$ & $3.12\pm0.002$ \\
    M-M27-G13 & III &$> 25$ & $3.9$ & $15\pm2$ & $0.20$ & $1.86\pm0.10$ & $3.11\pm0.06$ \\
    M-M27-G14 & III &$> 25$ & $3.9$ & $8\pm7$ & $0.20$ & $1.86\pm0.11$ & $3.03\pm0.03$ \\
    M-M27-G15 & III &$> 25$ & $3.8$ & $6\pm5$ & $0.20$ & $1.86\pm0.12$ & $2.90\pm0.07$ \\
    M-M27-G16 & III &$> 25$ & $3.8$ & $5\pm1$ & $0.20$ & $1.86\pm0.12$ & $2.93\pm0.01$ \\
    M-M27-G17 & III &$> 25$ & $3.8$ & $3\pm1$ & $0.23$ & $1.86\pm0.13$ & $2.91\pm0.01$ \\
    M-M27-G18 & III &$> 25$ & $3.7$ & $3\pm1$ & $0.23$ & $1.86\pm0.12$ & $2.84\pm0.06$ \\
    M-M27-G19 & III &$> 25$ & $3.7$ & $2.2\pm0.3$ & $0.27$ & $1.86\pm0.12$ & $2.82\pm0.04$ \\
    M-M27-G20 & III &$> 25$ & $3.6$ & $2.1\pm0.3$ & $0.28$ & $1.85\pm0.12$ & $2.83\pm0.01$ \\
    M-M29-G11 & II & $1.96\pm0.04$ & $-$ & $1\pm1$ & $0.38$ & $1.89\pm0.04$ & $-$ \\
    M-M29-G12 & II & $2.063\pm0.004$ & $-$ & $1.1\pm0.1$ & $0.37$ & $1.90\pm0.02$ & $-$ \\
    M-M29-G13 & II & $2\pm1$ & $-$ & $4\pm3$ & $0.28$ & $1.91\pm0.04$ & $-$ \\
    M-M29-G14 & III &$17\pm13$ & $5.2$ & $23\pm14$ & $0.25$ & $1.91\pm0.05$ & $3.39$\footnote{In this case the N102 model is missing and the N62 model has a short-lived MNS, so we do not put the error.} \\
    M-M29-G15 & III &$> 25$ & $4.8$ & $18\pm3$ & $0.23$ & $1.91\pm0.06$ & $3.30\pm0.01$ \\
    M-M29-G16 & III &$> 25$ & $4.7$ & $19\pm7$ & $0.22$ & $1.91\pm0.09$ & $3.25\pm0.03$ \\
    M-M29-G17 & III &$> 25$ & $4.6$ & $16\pm8$ & $0.22$ & $1.91\pm0.11$ & $3.17\pm0.07$ \\
    M-M29-G18 & III &$> 25$ & $4.5$ & $13\pm8$ & $0.23$ & $1.92\pm0.11$ & $3.12\pm0.07$ \\
    M-M29-G19 & III &$> 25$ & $4.6$ & $8\pm3$ & $0.25$ & $1.92\pm0.12$ & $3.21\pm0.17$ \\
    M-M29-G20 & III &$> 25$ & $4.8$ & $6\pm2$ & $0.26$ & $1.91\pm0.12$ & $3.09\pm0.12$ \\
    L-M27-G11 & I & $< 1$ & $-$ & $0.7\pm0.1$ & $0.40$ & $2.17\pm0.16$ & $-$ \\
    L-M27-G15 & II & $2.0\pm0.2$ & $-$ & $3\pm1$ & $0.35$ & $2.18\pm0.14$ & $-$ \\
    L-M27-G20 & III &$> 25$ & $5.7$ & $8\pm2$ & $0.25$ & $2.17\pm0.18$ & $3.48\pm0.07$ \\
    L-M29-G11 & I & $< 1$ & $-$ & $< 0.1$ & $0.20$ & $2.18\pm0.11$ & $-$ \\
    L-M29-G15 & I & $< 1$ & $-$ & $< 0.1$ & $0.20$ & $2.18\pm0.13$ & $-$ \\
    L-M29-G20 & I & $< 1$ & $-$ & $< 0.1$ & $0.15$ & $2.17\pm0.07$ & $-$ \\
    H-M27-G11 & III &$> 25$ & $2.7$ & $14\pm4$ & $0.26$ & $1.58\pm0.02$ & $2.48\pm0.01$ \\
    H-M27-G15 & III &$> 25$ & $2.5$ & $8\pm1$ & $0.20$ & $1.58\pm0.02$ & $2.34\pm0.06$ \\
    H-M27-G20 & III &$> 25$ & $2.5$ & $2\pm1$ & $0.21$ & $1.61\pm0.01$ & $2.30\pm0.02$ \\
    H-M29-G11 & III &$> 25$ & $3.2$ & $10\pm1$ & $0.28$ & $1.61\pm0.01$ & $2.46\pm0.17$ \\
    H-M29-G15 & III &$> 25$ & $3.0$ & $8\pm2$ & $0.21$ & $1.59\pm0.02$ & $2.48\pm0.05$ \\
    H-M29-G20 & III &$> 25$ & $2.8$ & $2.4\pm0.4$ & $0.24$ & $1.59\pm0.02$ & $2.39\pm0.04$ \\
    \hline
    \hline
  \end{tabular}
  \endgroup

\end{table*}

Table~\ref{tab:results} summarizes key postmerger observables obtained from our numerical simulations across varying EOS models, thermal indices $\Gamma_{\rm th}$, and total binary masses $m_0$.

\subsection{Overview of the merger process}
\label{sec:overview}

\begin{figure*}[htbp]
  \centering
    \includegraphics[width=0.49\textwidth]{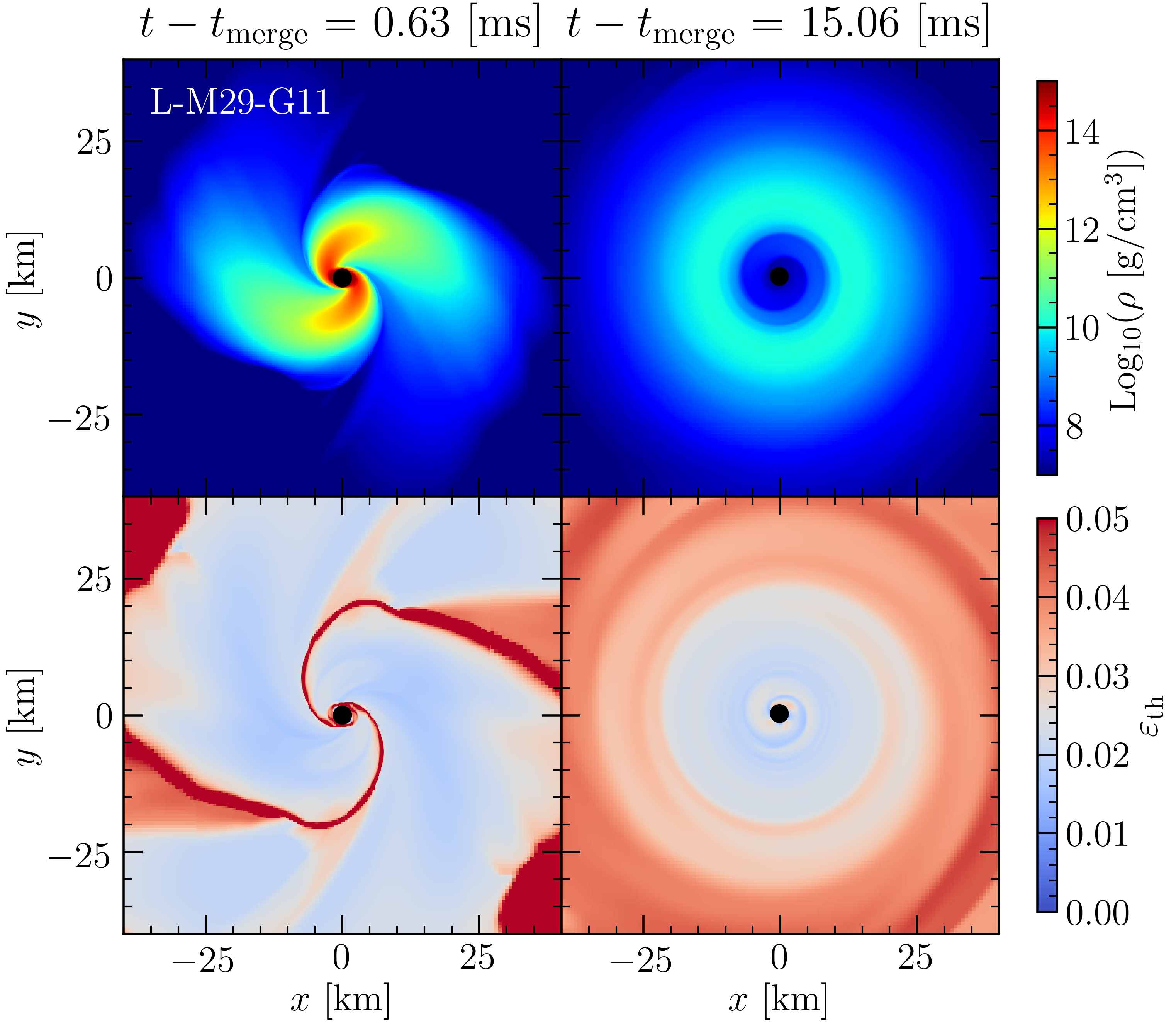}\hspace{0em}
    \includegraphics[width=0.49\textwidth]{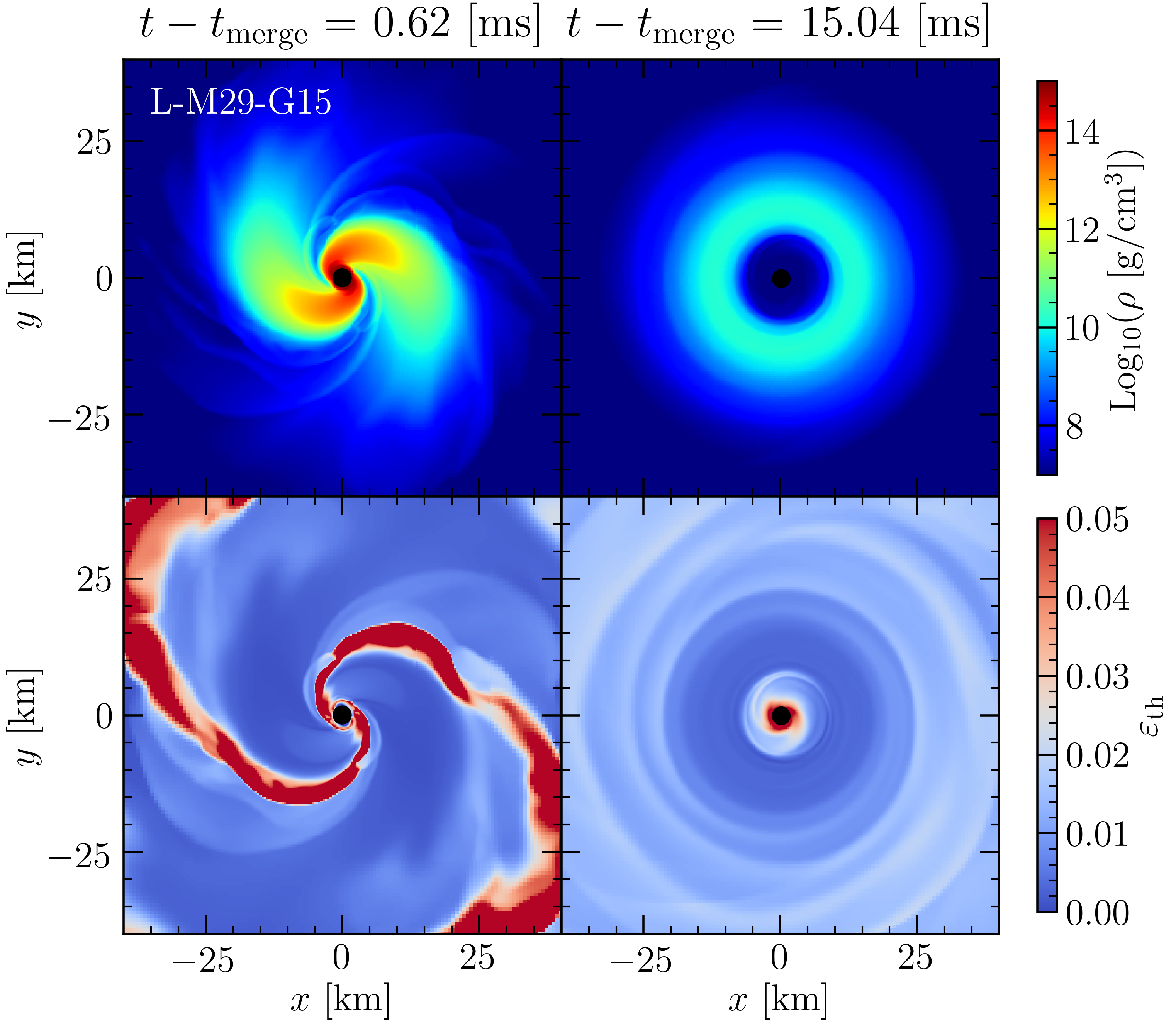}\\[0em]
    
    \includegraphics[width=0.49\textwidth]{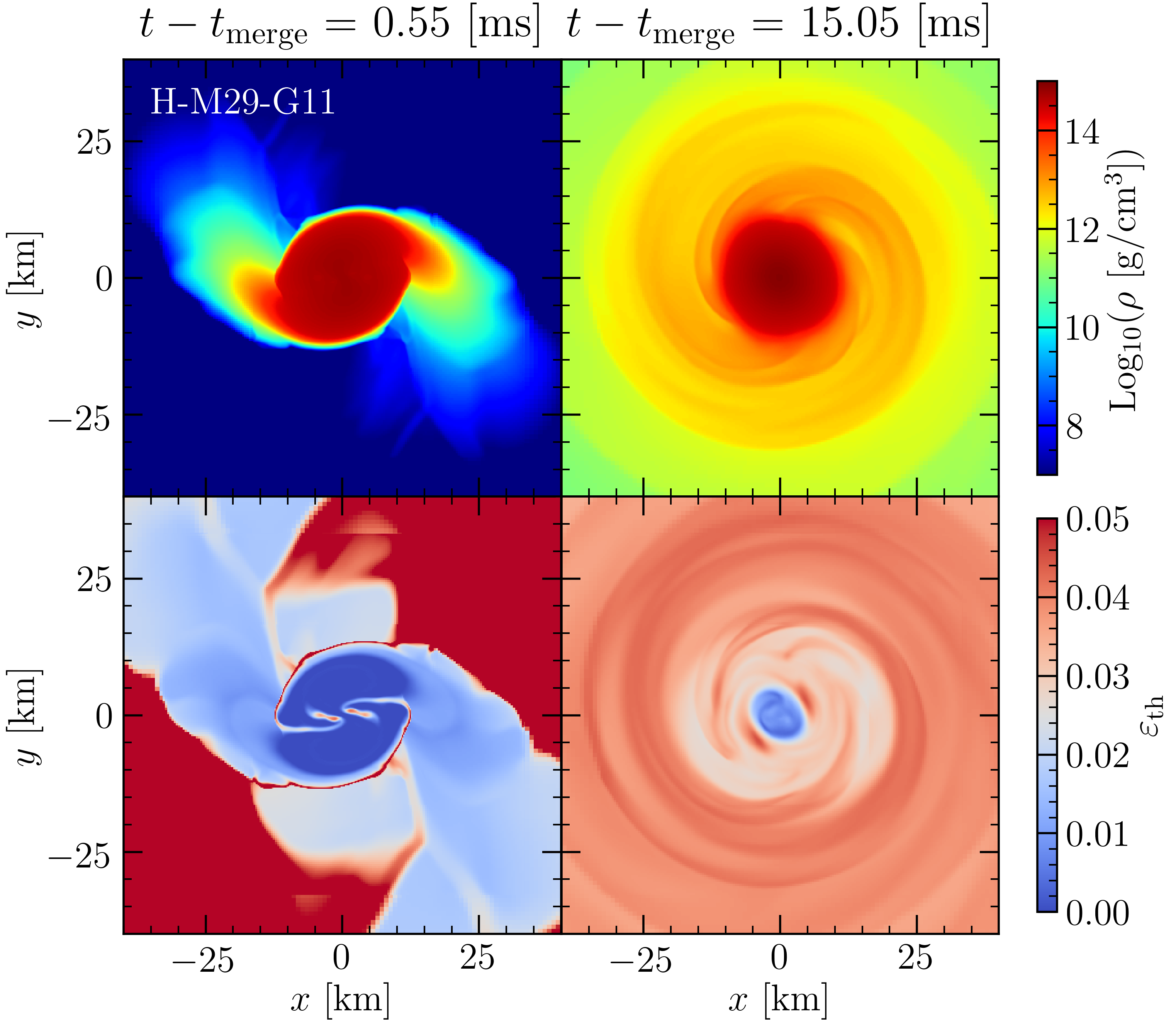}\hspace{0em}
    \includegraphics[width=0.49\textwidth]{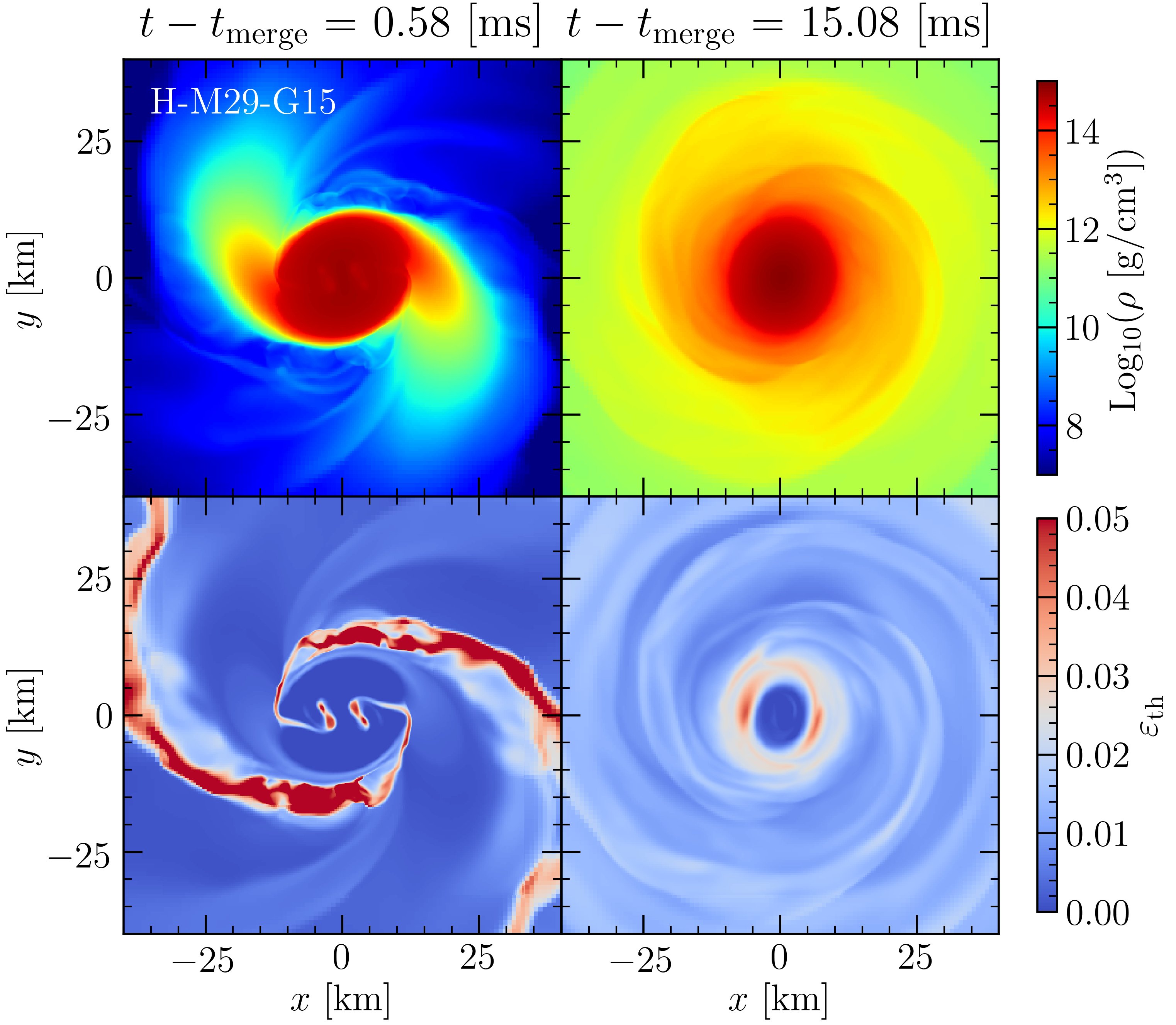}\\[0em]

  \caption{Snapshots of the density profile and the thermal part of the specific internal energy profile for several selected models on the $x$-$y$ plane. Four $2\times2$ subplots denote four different models, which are labeled on the upper left panel of each subplot (the meaning of the model can be found in Table~\ref{tab:models}). In each subplot, top/bottom rows display the density profiles/thermal part of the specific internal energy profiles. The columns stand for different time. For the models of the upper panels a BH (marked by a filled black circle) is formed after the merger while for the lower panels a MNS is formed.}
  \label{fig:dens_eth}
\end{figure*}

We first overview BNS merger processes, which have been extensively studied by many previous works (see, e.g., Ref.~\citep{Shibata:2019wef} for a review). 
Initially separated by approximately 45 km, the two NSs undergo inspiral driven by gravitational radiation reaction, ultimately leading to the eventual merger.
Following the merger, the fate of the remnant is governed primarily by the intricate interplay among gravitational attraction, pressure, and centrifugal forces arising from rapid rotation. 
Depending on the relative strengths of these forces, the postmerger object either collapses promptly into a BH or persists temporarily as a massive neutron star (MNS) before eventually collapsing or settling into a stable configuration.

Figure~\ref{fig:dens_eth} illustrates snapshots of the rest-mass density profile and the thermal part of the specific internal energy in the equatorial ($x$-$y$) plane for selected models, which can emphasize the influence of the thermal index $\Gamma_{\rm th}$ on both the formation of a BH accompanied by an accretion disk and the properties of the resultant MNS. 
In each $2\times2$ subplot, we present snapshots at two distinct time steps: shortly after the BNS merger ($t - t_{\rm merger} \sim 0.6$ ms) and when the merger remnant has stabilized ($t - t_{\rm merger} \sim 15$ ms). 
For the soft EOS scenarios illustrated in the upper two rows of Fig.~\ref{fig:dens_eth}, prompt gravitational collapse into a BH occurs rapidly due to the high total binary mass. 
As surrounding neutron-rich matter continuously accretes onto the newly formed BH, an accretion disk system emerges, clearly visible in the snapshots. 
Conversely, the enhanced pressure effectively prevents immediate collapse in the stiff EOS scenarios depicted in the lower two rows. 
Consequently, a MNS forms and remains stable throughout the simulation period, distinctly demonstrating the EOS-dependent outcomes of the merger events.

\subsection{Classification of the merger remnant}

\begin{figure}[!htbp]
  \centering
    \includegraphics[width=0.48\textwidth]{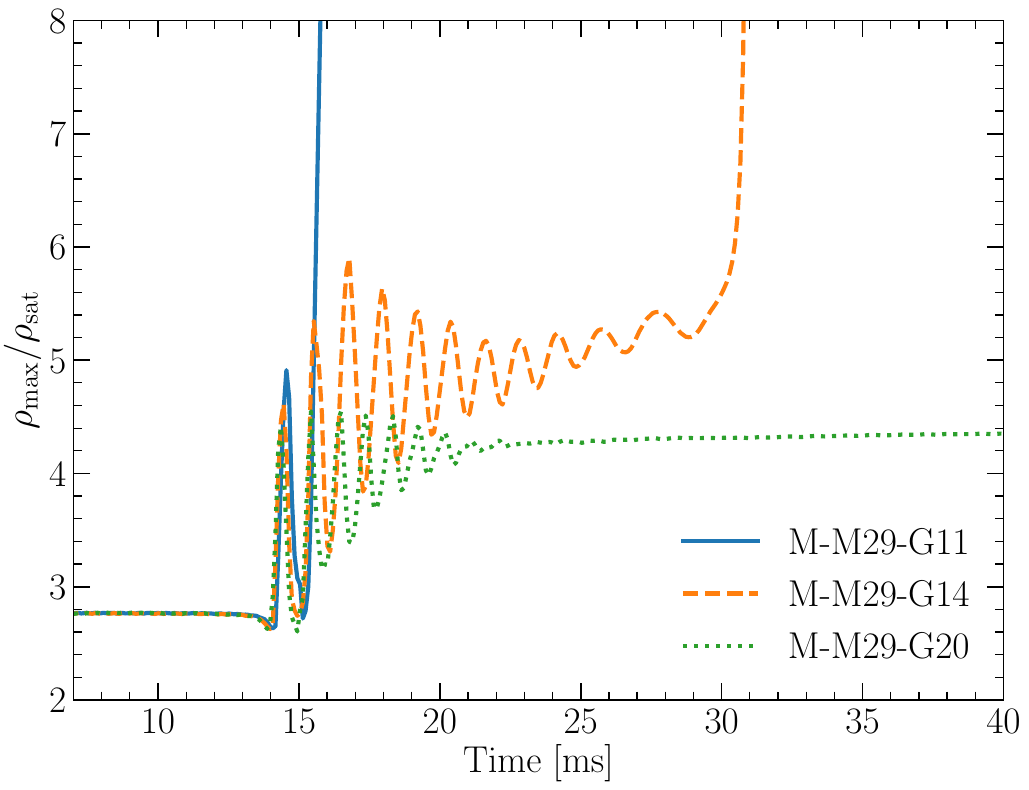}

  \caption{Evolution of the maximum rest-mass density $\rho_{\rm max}$. Three models with different types of remnant are shown: Nearly prompt collapse, delayed collapse, and the long-lived NS case.}
  \label{fig:rho_max}
\end{figure}

We follow the classification scheme proposed in Ref.~\citep{Hotokezaka:2011dh} to categorize merger remnants into three distinct types based on the lifetime of the MNS following the merger:
\begin{itemize}
    \item Type I: Prompt collapse, characterized by immediate formation of a BH;
    \item Type II: Short-lived MNS, surviving for a brief interval ($\tau_{H}< 5$ ms) before collapsing;
    \item Type III: Long-lived MNS, persisting beyond 5 ms after the merger ($\tau_{H}> 5$ ms).
\end{itemize}
Here, the MNS lifetime is defined as $\tau_{H} = t_{\rm collapse} - t_{\rm merger}$, where $t_{\rm merger}$ is the time at which the GW amplitude becomes maximal.

Figure~\ref{fig:rho_max} presents the evolution of the maximum rest-mass density, $\rho_{\rm max}$, for representative models illustrating different remnant types: M-M29-G11 (type II), M-M29-G14 (type III), and M-M29-G20 (type III). 
For the nearly prompt collapse scenario represented by model M-M29-G11, the density $\rho_{\rm max}$ experiences an initial transient oscillation following the merger, and then rapidly and monotonically increases, resulting in the immediate formation of a BH. 
Model M-M29-G14 exhibits distinct oscillatory behavior of $\rho_{\rm max}$, with each oscillation accompanied by a systematic increase in maximum density until a delayed collapse occurs. 
In contrast, the long-lived MNS scenario, exemplified by model M-M29-G20, features initial oscillations that gradually damp, and the system settles into a stable MNS, maintaining a nearly constant maximum density throughout the simulation duration. 
These diverse outcomes underscore the significant influence of the thermal index $\Gamma_{\rm th}$ on the merger remnant characteristics. 
Specifically, smaller values of $\Gamma_{\rm th}$ yield lower thermal pressures, resulting in a denser core and a correspondingly higher probability of collapse. 
This sensitivity is particularly pronounced when the system resides near the threshold separating collapse and stability.

\begin{figure}[t]
  \centering
    \includegraphics[width=0.48\textwidth]{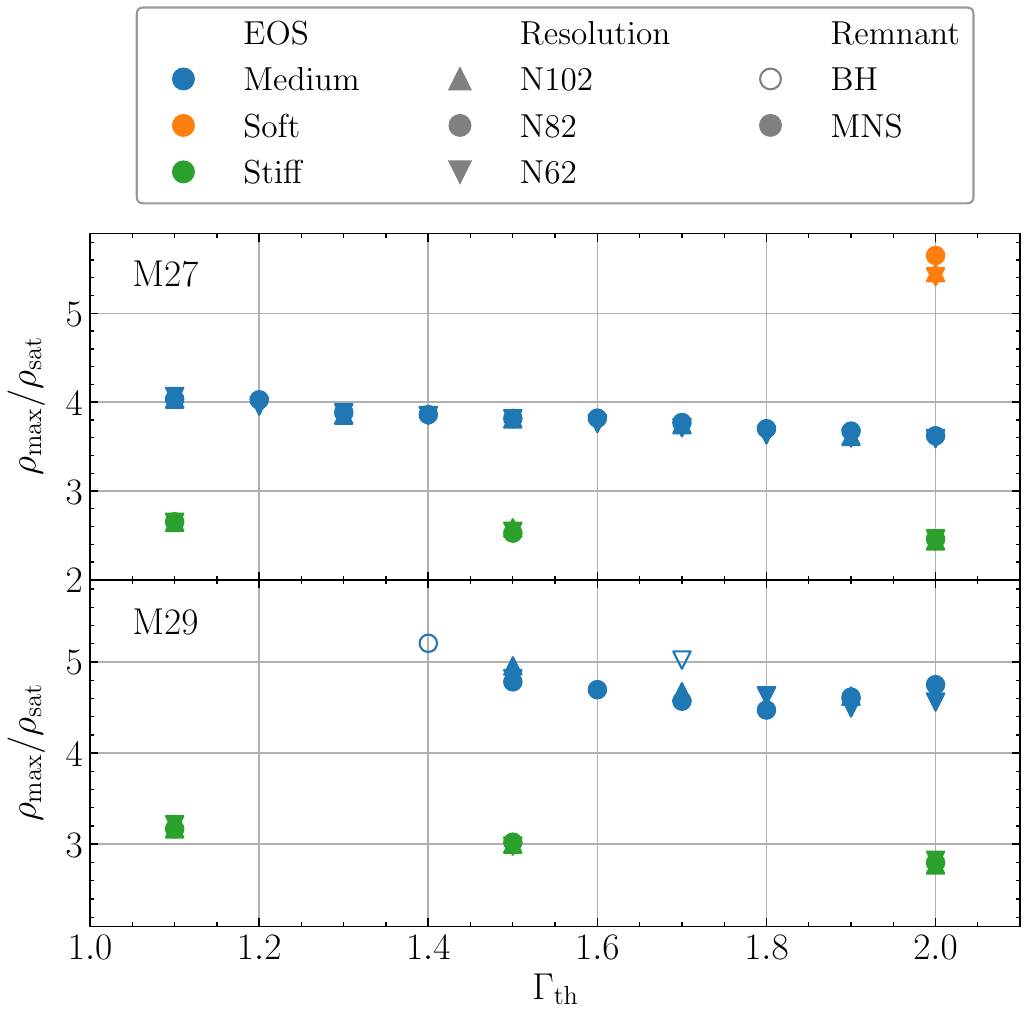}

  \caption{Maximum rest-mass density $\rho_{\rm max}$ at 15 ms after the onset of merger for different models. The top panel shows the results for models with the total mass of $2.7M_\odot$, while the bottom panel shows the results for models with the total mass of $2.9M_\odot$. The scatters with different colors denote the different EOSs, i.e., blue for medium, orange for soft, and green for stiff. The up triangle/dot/down triangle are the results with N102/N82/N62 resolutions, respectively. The filled/unfilled symbols denote the type of the remnant (MNS/BH) at the end of the simulation, respectively.}
  \label{fig:rho_max_scatters}
\end{figure}

Figure~\ref{fig:rho_max_scatters} summarizes the maximum rest-mass density $\rho_{\rm max}$ evaluated at approximately 15 ms after the onset of merger across various simulation scenarios. 
It is evident that, for fixed EOS and total mass, increasing $\Gamma_{\rm th}$ tends to slightly reduce the maximum density, consistent with observations in Fig.~\ref{fig:rho_max}. 
Notably, among the soft EOS models, only L-M27-G20 (total mass $2.7M_\odot$) maintains a stable MNS at this time point. 
For medium EOS models with total mass $2.7~M_\odot$, stable remnants persist throughout the simulation irrespective of the chosen thermal index. 
However, at higher total mass ($2.9~M_\odot$), medium EOS models with $\Gamma_{\rm th} \leq 1.4$ collapse into BHs, whereas those with $\Gamma_{\rm th} \geq 1.5$ sustain long-lived NSs. 
Conversely, all stiff EOS scenarios consistently form stable, long-lived NSs regardless of variations in total mass and thermal index. 

Convergence studies, illustrated in Fig.~\ref{fig:rho_max_scatters} and indicated by distinct markers for different resolutions (N62, N82, and N102), demonstrate robust numerical convergence for most scenarios. 
Nevertheless, discrepancies arise for $M=2.9\,M_\odot$ at intermediate $\Gamma_{\rm th}$ values (1.7), where simulations at lower resolution (N62) erroneously predict delayed collapse, whereas higher resolution results (N82 and N102) consistently exhibit noncollapse outcomes, highlighting the necessity of appropriate numerical resolution in accurately determining the remnant classification.

\subsection{The dynamics of the merger process}

We further describe the details of the merger process for the different models presented in Fig.~\ref{fig:dens_eth}.
For the cases of a soft EOS (shown in the upper two rows of Fig.~\ref{fig:dens_eth}), prompt collapse to a BH occurs swiftly after the onset of the merger due to the high total mass of the system. 
Under these conditions, most of the NS matter quickly falls into the newly formed BH, resulting in very limited ejecta and disk mass. 
Nevertheless, the relative scarcity of surrounding matter allows clear visualization of the resulting accretion disk structure. 
As illustrated in the top row of Fig.~\ref{fig:dens_eth}, by approximately 0.6 ms postmerger, the BH is formed, and the residual matter continuously spirals inward, manifesting as pronounced two-armed spiral patterns. 
These spiral arms exhibit sensitivity to the choice of the thermal index; specifically, larger values of $\Gamma_{\rm th}$ result in significantly wider and denser spiral arms compared to the scenario with a lower thermal index. 
This occurs because a larger value of $\Gamma_{\rm th}$ generates greater thermal pressure, driving a broader spatial distribution of matter around the BH.

By approximately 15 ms after the merger, the accretion process stabilizes, resulting in comparable accretion disk structures for both the G11 and G15 models. 
Additionally, the second row of Fig.~\ref{fig:dens_eth} illustrates the thermal component of the internal energy, highlighting that a larger value of $\Gamma_{\rm th}$ enhances the shock heating efficiency. 
Consequently, the heated matter expands to larger radii, reducing its localized thermal energy density and making $\varepsilon_{\rm th}$ appear diminished in the snapshots.

In contrast, for stiff EOS models (bottom two rows of Fig.~\ref{fig:dens_eth}), the dynamics differ considerably. 
Here, even the lowest thermal pressure scenario (G11 model) generates sufficient internal pressure and centrifugal support to prevent immediate gravitational collapse, thereby facilitating significant mass ejection following the merger event. 
Due to substantial shock heating, the remnant matter expands extensively away from the central region. 
As expected, this expansion becomes increasingly pronounced with higher values of $\Gamma_{\rm th}$, corresponding to elevated thermal pressures. 
The third row of Fig.~\ref{fig:dens_eth} clearly demonstrates the evolution of the rest-mass density for scenarios with both low and high values of $\Gamma_{\rm th}$. 
For the model with smaller thermal index (left two panels), at about 0.6 ms after the onset of the merger, a distinct and relatively compact spiral arm structure appears around the central object. 
Conversely, models with a higher thermal index (right two panels) display immediate and widespread matter dispersion post-merger, resulting in a more diffuse and extended material distribution around the central object.

By approximately 15 ms after the merger, the MNS reaches a quasisteady configuration. A higher thermal index systematically leads to increased expansion of the remnant, thereby producing a less compact MNS structure compared to the lower thermal index case. Regarding the thermal component of the specific internal energy (bottom row in Fig.~\ref{fig:dens_eth}), we observe analogous behavior to the soft EOS scenarios: higher $\Gamma_{\rm th}$ values lead to broader material dispersion, consequently lowering the localized values of $\varepsilon_{\rm th}$ visible in the snapshots. 

\subsection{Dynamical mass ejection}

\begin{figure}[t]
  \centering
    \includegraphics[width=0.48\textwidth]{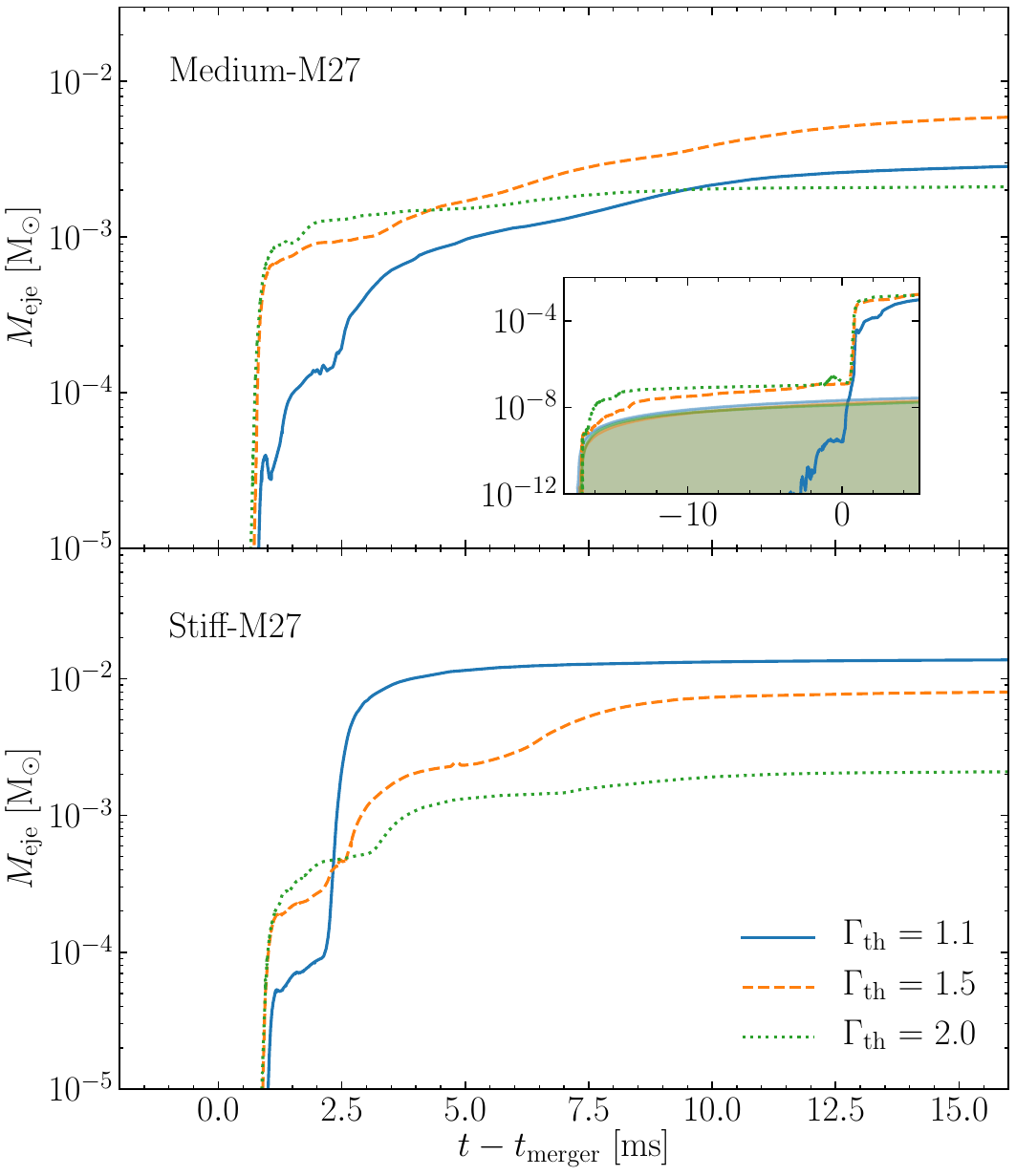}
  \caption{Evolution of the dynamical ejecta mass for the models with medium (upper panel) and stiff (lower panel) EOS. The total mass of these models is all $2.7~M_\odot$. The blue solid lines denote the result with $\Gamma_{\rm th}=1.1$, the orange dashed lines with $\Gamma_{\rm th}=1.5$, and the green dotted lines with $\Gamma_{\rm th}=2.0$. The color-shaded regions in the inset plots of the upper panel denote the baryonic mass conservation errors for the corresponding models.}
  \label{fig:ejecta}
\end{figure}

During BNS mergers, neutron-rich matter can be dynamically ejected on short timescales, typically $\lesssim$10 ms. 
This phenomenon, known as dynamical mass ejection, plays a critical role in shaping observable electromagnetic counterparts, such as kilonovae, and is comprehensively reviewed in Ref.~\citep{Shibata:2019wef}. 
The dynamical ejecta originate primarily from two distinct physical mechanisms: shock heating and angular momentum transport via torque exerted by the nonaxisymmetric rotating remnant.

Immediately following the onset of the merger, shock waves generated at the interface of the colliding NSs induce substantial heating of matter. 
The resulting thermal expansion pushes this heated matter outward, and the magnitude of this outward thrust is strongly influenced by the thermal index $\Gamma_{\rm th}$ employed in the simulations. 
Specifically, Fig.~\ref{fig:ejecta} illustrates the temporal evolution of dynamically ejected mass for binary systems with the total mass of $2.7~M_\odot$, employing medium and stiff EOSs with various $\Gamma_{\rm th}$ values. 
We observe that, within approximately 2 ms after the onset of the merger, a clear trend emerges: higher values of $\Gamma_{\rm th}$ consistently produce more substantial dynamical ejecta masses for both medium (upper panel) and stiff (lower panel) EOS scenarios. 
This outcome is directly attributed to the dependence of shock heating efficiency on $\Gamma_{\rm th}$, as expressed in Eq.~\eqref{eq:p_th}: 
Higher values of $\Gamma_{\rm th}$ lead to greater thermal pressure, enhancing outward expansion and thereby facilitating easier mass ejection.

In subsequent stages, angular momentum transport driven by torques from the nonaxisymmetric MNS becomes increasingly significant. 
The efficiency of this angular momentum transport depends critically on the remnant's compactness. 
With lower thermal pressure corresponding to lower values of $\Gamma_{\rm th}$, the central MNS tends to achieve greater compactness, thus intensifying the torque-driven ejection process. 
The lower panel of Fig.~\ref{fig:ejecta} clearly exemplifies this effect for stiff EOS simulations, where dynamical ejecta mass increases notably with lower $\Gamma_{\rm th}$ values beyond 2 ms postmerger.

In contrast, the upper panel of Fig.~\ref{fig:ejecta} reveals a more complex behavior for medium EOS models. 
Here, the largest ejecta mass is found at an intermediate thermal index ($\Gamma_{\rm th}=1.50$), rather than at the highest value. 
This divergence arises due to the ``trapping effect" as detailed in Ref.~\citep{Hotokezaka:2012ze}. 
For softer EOS models compared to stiffer ones, the resulting MNS remnant is inherently more compact, causing potential ejecta material to reside deeper within the gravitational potential well. 
As a consequence, escaping from the gravitational field requires greater kinetic energy, making mass ejection more challenging. 
Additionally, a smaller $\Gamma_{\rm th}$ value exacerbates this effect, as the MNS is more compact. 
This competition between torque-driven ejection and gravitational trapping is particularly pronounced in medium EOS scenarios at lower thermal indices.

Besides, we also show the baryonic mass conservation error in the upper panel of Fig.~\ref{fig:ejecta}. 
We can see that the overall errors are below $\sim~10^{-8}~M_\odot$, which implies that the baryonic mass is well conserved during the simulation for the present purpose. 
In the inspiral phase ($t - t_{\rm merger} \lesssim 0$ ms), the ``spurious mass ejection'' for the model with $\Gamma_{\rm th}=1.5~\text{and}~2.0$ lies above the conservation error lines, while that of the model with $\Gamma_{\rm}=1.1$ can be neglected compared with the conservation error. 
This may indicate that for the inspiral phase, the lower $\Gamma_{\rm th}$ value is a better choice for physical modeling. This suggests that a nonconstant $\Gamma_{\rm th}$ hybrid framework needs to be considered in future work.
In addition, the criteria $-u_t>1$ of the ejected matter may also affect the above issue.

\begin{figure}[t]
  \centering
    \includegraphics[width=0.45\textwidth]{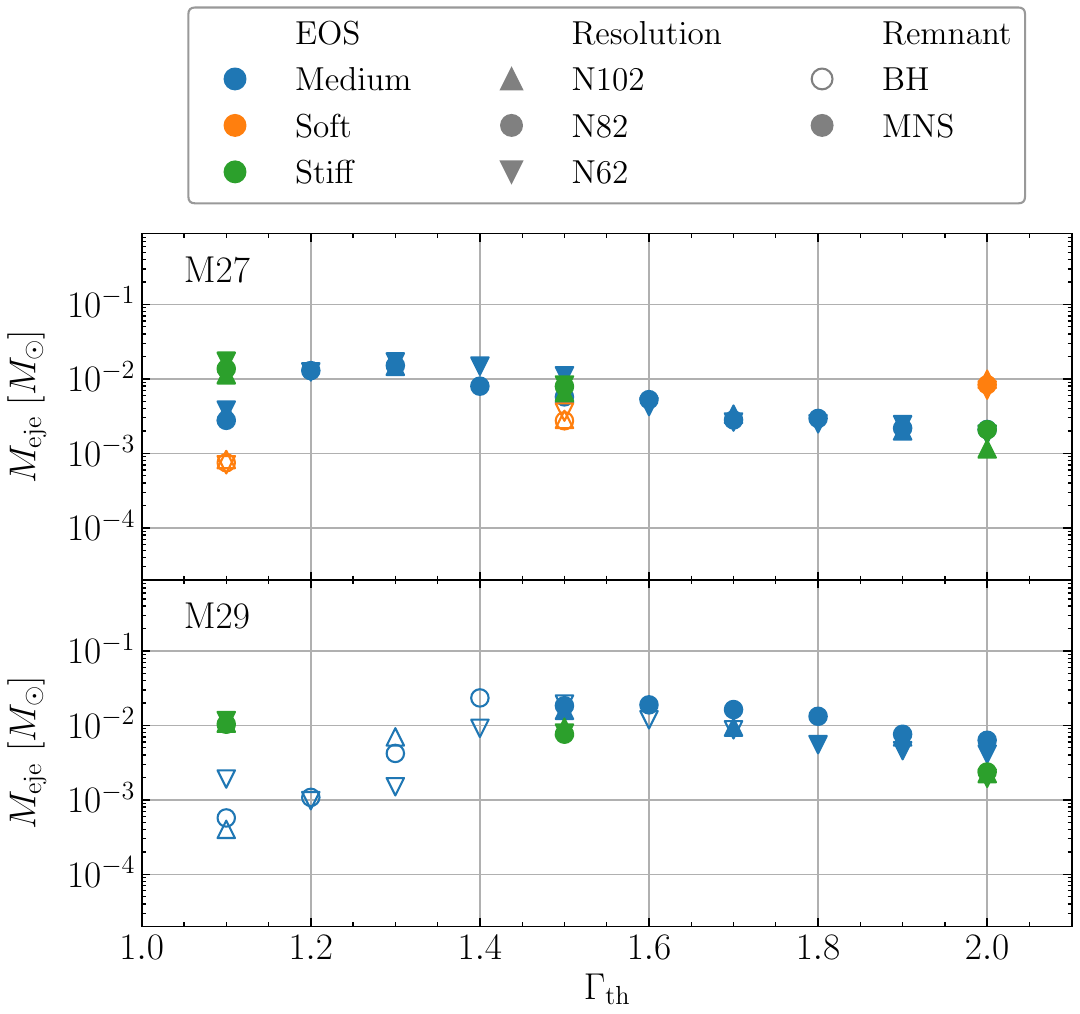}

  \caption{Dynamical ejecta mass at $\sim$ 15 ms after the onset of the merger. The meaning of the legend is the same as in Fig.~\ref{fig:rho_max_scatters}.}
  \label{fig:ejecta_scatters}
\end{figure}

Figure~\ref{fig:ejecta_scatters} illustrates the dependence of dynamical ejecta mass evaluated at 15 ms after the onset of the merger on the thermal index $\Gamma_{\rm th}$. 
The consistency across different numerical resolutions, as evident from the figure, indicates fair convergence (within a factor of $\sim 2$ for N102) for the presented results. 
Note that certain models are absent from the plot due to prompt BH formation, leaving negligible or no ejecta. 
We note that for the result with a low resolution like N62 the dynamical ejecta mass is likely to have a large error.

As depicted in the upper panel, the variation of ejecta mass with $\Gamma_{\rm th}$ exhibits notable EOS-dependent behavior. 
For models utilizing a soft EOS, the ejecta mass consistently increases with higher values of $\Gamma_{\rm th}$. 
This occurs primarily because the soft EOS scenarios predominantly result in immediate BH formation; thus, significant ejecta mass production is limited to the highest thermal index scenario ($\Gamma_{\rm th}=2.0$), where the remnant is a long-lived MNS, enhancing dynamical mass ejection through sustained thermal pressure. 
For the remaining soft EOS cases with lower values of $\Gamma_{\rm th}$, ejecta formation is solely attributed to short-term shock heating. 
Consequently, increased thermal pressure from a higher value of $\Gamma_{\rm th}$ directly translates into more substantial ejecta.

Conversely, for stiff EOS models, an inverse trend is observed: the ejecta mass monotonically decreases as the increase of $\Gamma_{\rm th}$. 
This phenomenon arises from the torque-driven mass ejection mechanism, which becomes predominant after initial shock-driven ejection. 
In scenarios with a stiff EOS, a smaller thermal index produces a more compact remnant MNS, thereby amplifying the efficiency of angular momentum transport to the surrounding matter. 
This augmented torque mechanism results in greater kinetic energy imparted to the ejecta, consequently enhancing the dynamical ejecta mass.

For the medium EOS, the dynamical ejecta mass exhibits a nonmonotonic behavior, initially increasing with $\Gamma_{\rm th}$, reaching a maximum around $\Gamma_{\rm th} \approx 1.3$, and subsequently decreasing for higher values. 
This characteristic behavior can be attributed to the competing influences of gravitational trapping and torque-driven mass ejection.
Specifically, for low $\Gamma_{\rm th}$ values ($\Gamma_{\rm th}\lesssim 1.3$), gravitational trapping due to the compactness of the central MNS is dominant, requiring greater kinetic energy for matter to overcome gravitational potential energy and escape. 
Hence, as thermal pressure rises, more ejecta are initially produced. 
However, beyond the critical value ($\Gamma_{\rm th}\approx 1.3$), further increases in thermal pressure result in a less compact central object, reducing the efficiency of torque-driven mass ejection and ultimately lowering the ejecta mass.

The lower panel of Fig.~\ref{fig:ejecta_scatters} details similar analyses for systems with the higher total mass of $2.9~M_\odot$. 
For soft EOS scenarios at this higher total mass, all considered cases promptly collapse to BHs, yielding negligible ejecta masses ($\ll 10^{-4}M_\odot$) and thus not depicted. 
For stiff EOS models, the inverse relationship between $\Gamma_{\rm th}$ and ejecta mass remains consistent with the lower mass case. 
However, for medium EOS models at this increased mass, the peak in ejecta mass distribution shifts to a larger $\Gamma_{\rm th}$ value than for the $m_0=2.7~M_\odot$ case. 
This shift occurs because, for more compact remnants at higher total masses, the gravitational trapping effect becomes increasingly significant at lower thermal pressures, thereby shifting the point at which torque-driven mechanisms dominate to higher $\Gamma_{\rm th}$ values. 
Consequently, the peak ejecta mass emerges at an elevated thermal index, highlighting the complex interplay between thermal effects and gravitational dynamics in BNS mergers.

\subsection{Gravitational waveform and spectra}

\begin{figure*}[!htbp]
    \centering
    
    \includegraphics[width=0.49\textwidth]{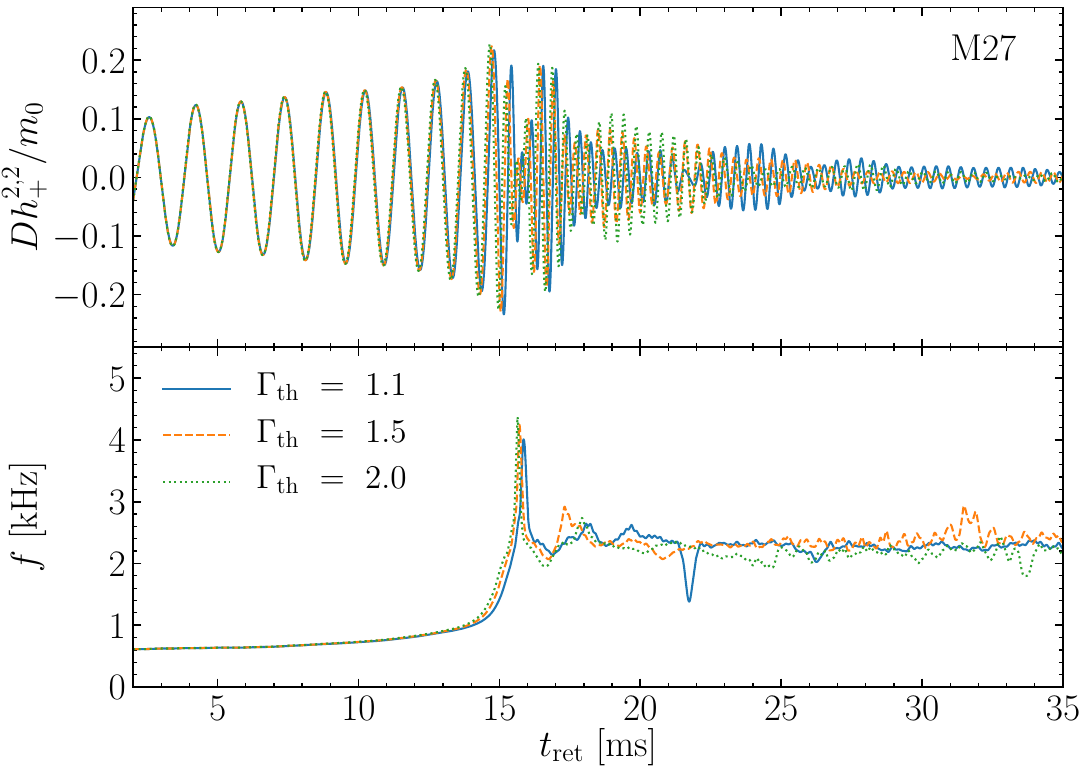}\hspace{0em}
    \includegraphics[width=0.49\textwidth]{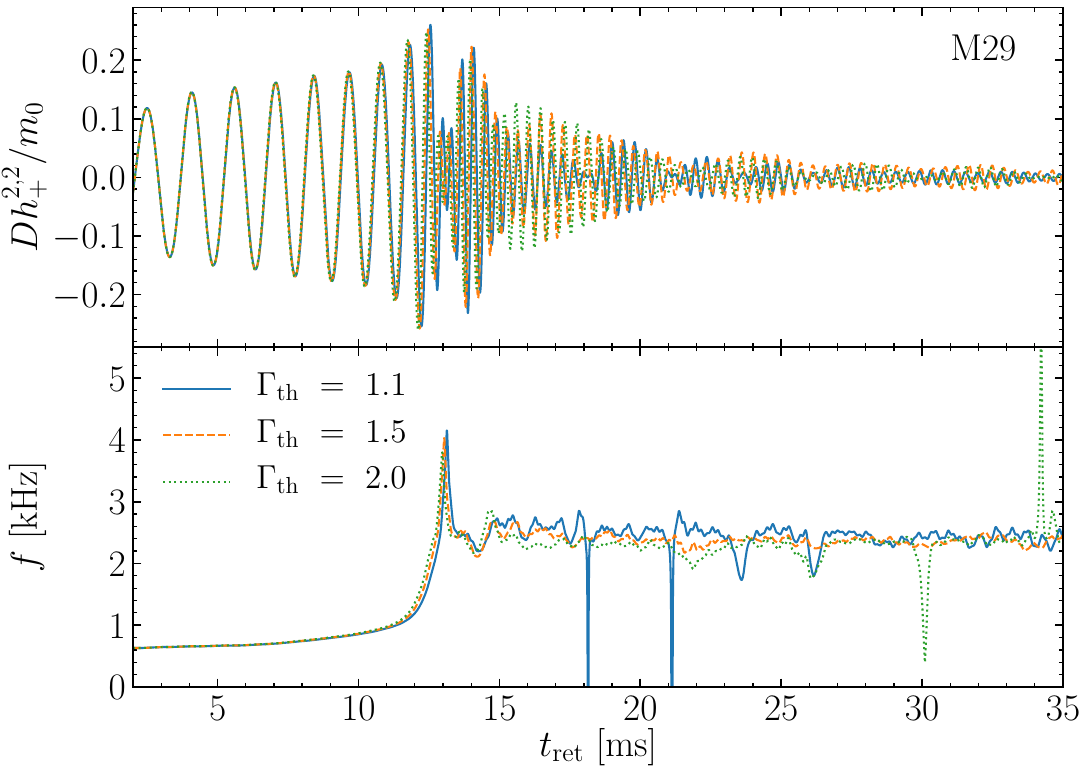}\\[0em]
    
    \caption{Gravitational waveforms ({\it upper}) and instantaneous GW frequencies ({\it lower}) of models with stiff EOS. The left subplot shows the results for $m_0 = 2.7\,M_\odot$, while the right subplot shows the results for $m_0 = 2.9\,M_\odot$. Models with different $\Gamma_{\rm}$ values are plotted with different line colors and styles.}
    \label{fig:gw_fgw}
\end{figure*}

Figure~\ref{fig:gw_fgw} displays gravitational waveforms and corresponding instantaneous frequencies for models employing a stiff EOS across different thermal indices, $\Gamma_{\rm th}=1.1, 1.5$, and $2.0$. 
The left panel illustrates results for $m_0 = 2.7M_\odot$, while the right panel for $m_0 = 2.9M_\odot$. 
During the inspiral phase, waveforms remain nearly indistinguishable across the range of $\Gamma_{\mathrm{th}}$ values examined since thermal heating is negligible.
Immediately after the onset of the merger, pronounced heating due to shock interactions substantially elevates the temperature of the merger remnant, thus amplifying thermal effects. 
Consequently, the gravitational waveforms in this postmerger regime exhibit clearly distinguishable characteristics in amplitude, oscillation frequency, and damping timescale for different choices of $\Gamma_{\rm th}$. 

\begin{figure*}[!htbp]
    \centering
    \includegraphics[width=0.95\textwidth]{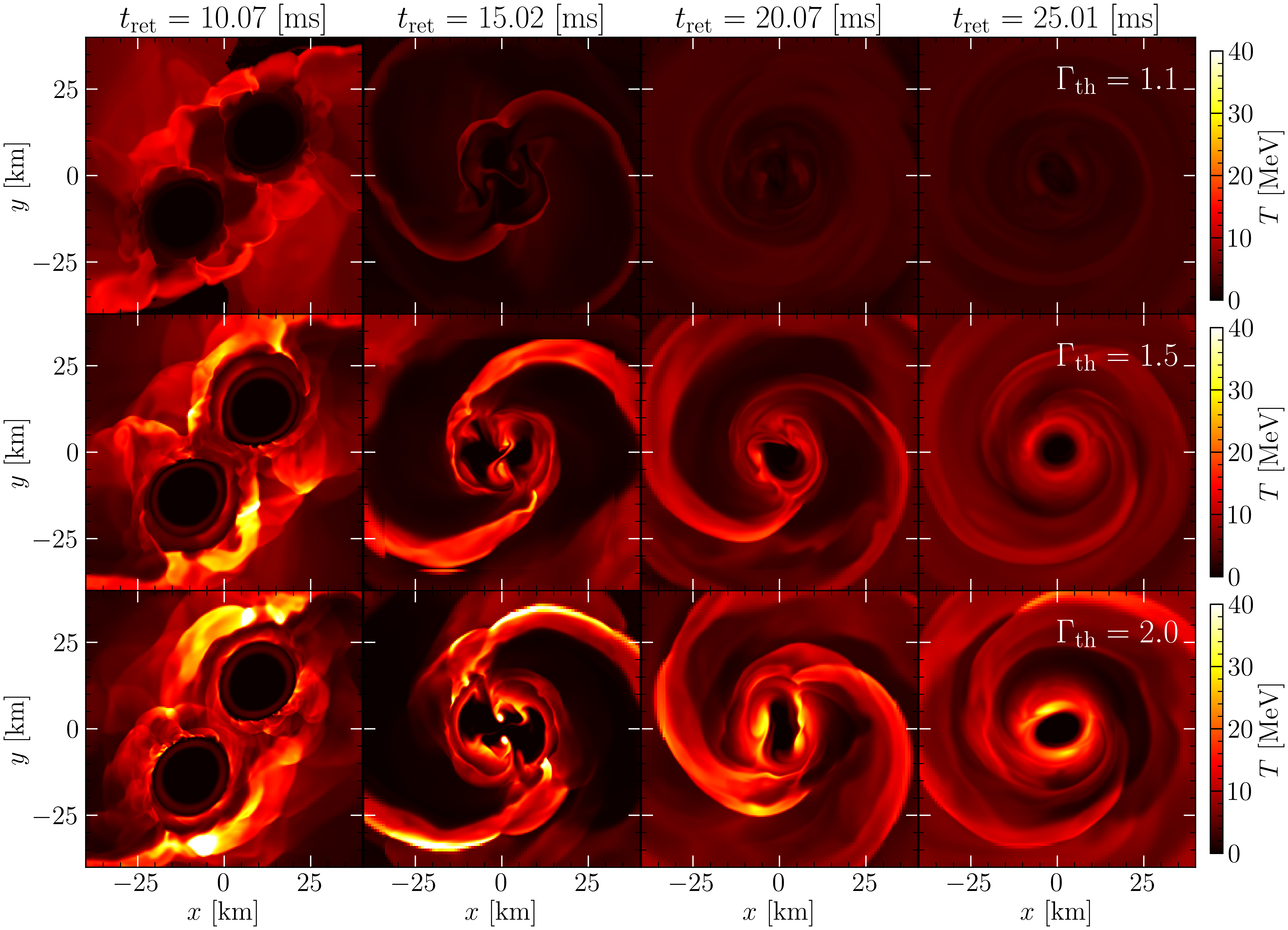}
    \caption{Snapshots of the temperature profile for several selected models on the $x$-$y$ plane. All the selected models are with stiff EOS and $m_0 = 2.7\,M_\odot$. The first/second/third rows stand for models with $\Gamma_{\rm th}=1.1/1.5/2.0$, and different columns are results evaluated at different retarded times.}
    \label{fig:T_profile}
\end{figure*}

Figure~\ref{fig:T_profile} contrasts the $x$-$y$ plane temperature profiles for the $\Gamma_{\rm th}$ = 1.1, 1.5, and 2.0 models at four representative retarded times ($t_{\rm ret} \approx$ 10, 15, 20, 25 ms), thereby providing direct thermal context for the gradual divergence of the gravitational-wave signals highlighted in Fig.~\ref{fig:gw_fgw}.
\begin{itemize}
    \item Pre-merger ($t_{\rm ret}\approx$ 10 ms). Apart from thin shock-heated surface layers ($T \lesssim$ 10 MeV), the stars remain essentially cold and the thermal pressure contribution is dynamically negligible, consistent with the nearly identical GW phase evolution up to this epoch (see Fig.~\ref{fig:gw_fgw}).
  \item Merger onset ($t_{\rm ret}\approx$ 15 ms). The collision interface is swept by spiral shocks that raise local temperatures to $T \gtrsim$ 20 MeV ($T \gtrsim$ 10 MeV for $\Gamma_{\rm th}$ = 1.1).  The radial and azimuthal extent of the hot matter already depends on $\Gamma_{\rm th}$: larger $\Gamma_{\rm th}$ produces broader, more asymmetric high-$T$ regions, marking the moment at which the instantaneous GW frequency curves begin to separate.
  \item Early remnant stage ($t_{\rm ret}\approx$ 20 ms). A differentially rotating remnant forms. For $\Gamma_{\rm th}$ = 2.0 the shock-heated material envelops a large fraction of the shear layer, whereas for $\Gamma_{\rm th}$ = 1.1 it is confined to a slender ring. The enhanced thermal pressure in the higher-$\Gamma_{\rm th}$ cases damps the core’s quadrupole oscillations more efficiently, lowering the plateau of the post-merger GW frequency by $\sim$ 0.2--0.3 kHz.
  \item Quasi-stationary phase ($t_{\rm ret}\approx$ 25 ms). In the $\Gamma_{\rm th}$ = 1.1 model the envelope has cooled to $T \lesssim$ 10 MeV, while for $\Gamma_{\rm th} \geq $ 1.5 temperatures of 15--25 MeV persist in the spiral-arm region. The resulting extra pressure support explains the $\sim$ 10 \% amplitude and frequency differences that remain visible in the late-time GW signal.
\end{itemize}

\begin{figure*}[htbp]
    \centering
    
    \includegraphics[width=0.45\textwidth]{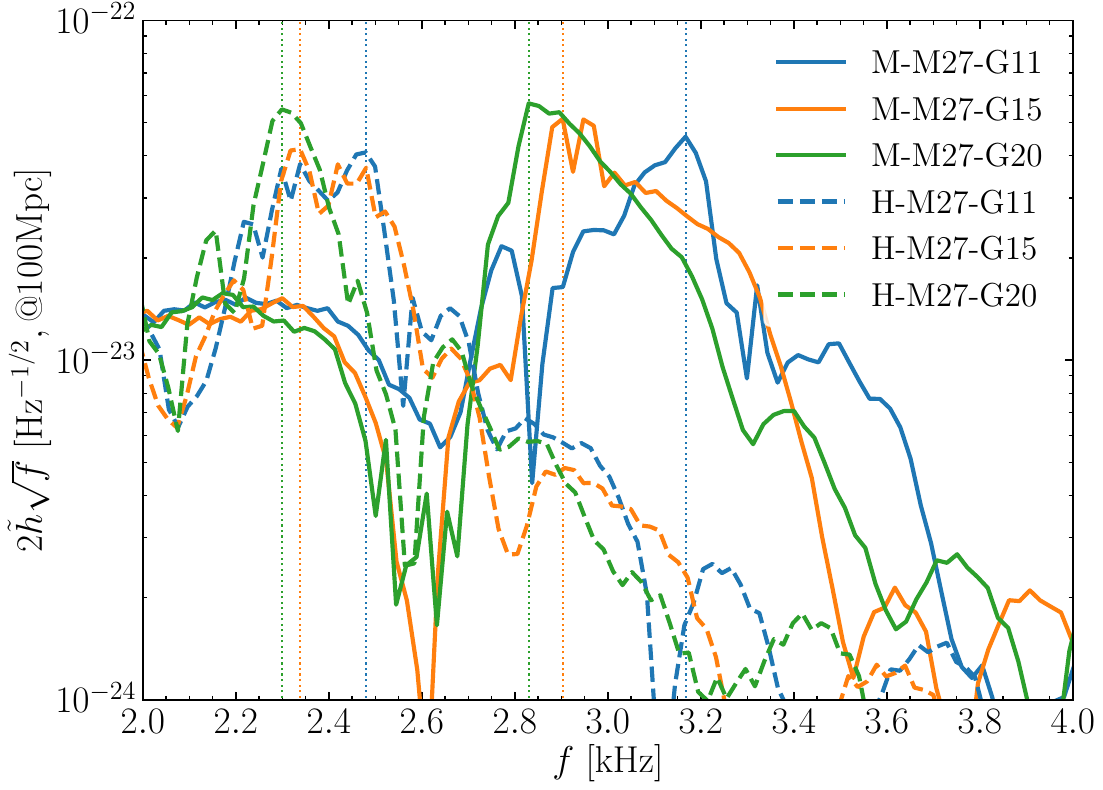}\hspace{0em}
    \includegraphics[width=0.45\textwidth]{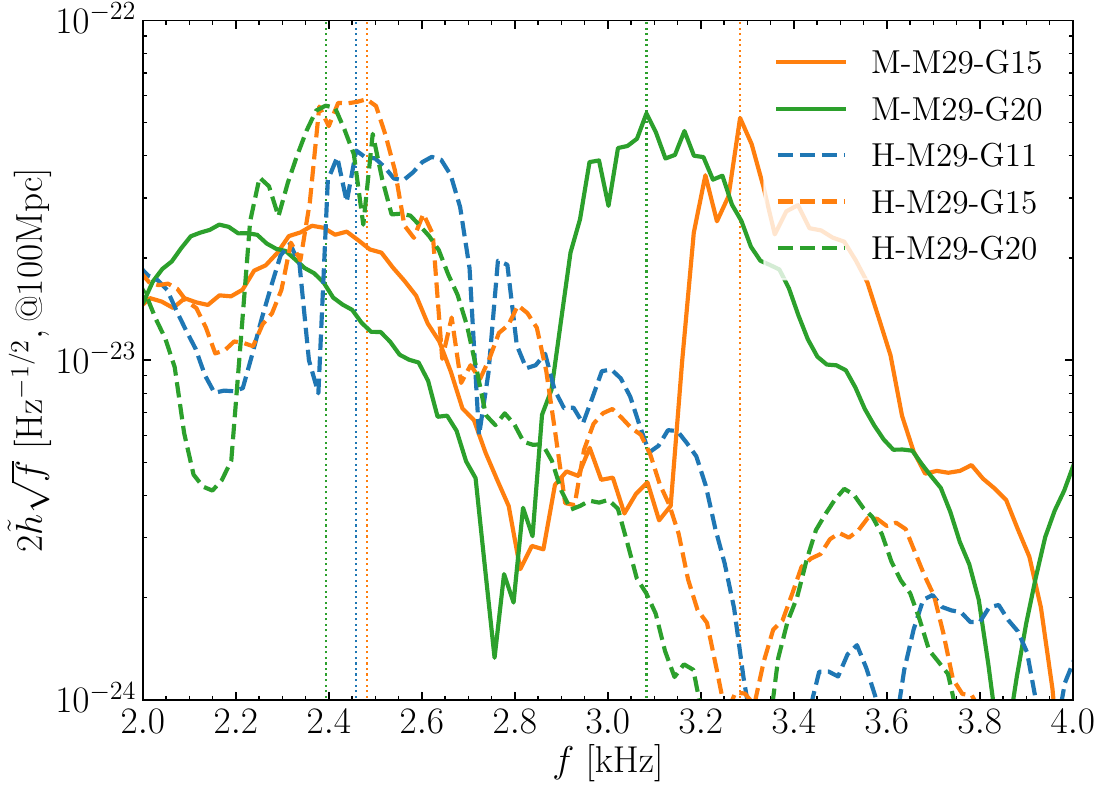}

    \caption{ASD $2\tilde{h}\sqrt{f}~{\rm [Hz^{-1/2}]}$ of models with different EOSs and $\Gamma_{\rm th}$ at a distance of 100 Mpc. Solid lines are the models with medium EOS, while dash-dot lines are those with stiff EOS. Vertical dashed lines denote the frequency of the dominant peak $f_2$. The left and right panels correspond to $m_0 = 2.7~M_\odot$ and $m_0 = 2.9~M_\odot$, respectively. Note that the prompt collapse models are excluded on this plot.}
    \label{fig:gw_sepctra}
\end{figure*}

We further perform Fourier transformations of the time-domain waveforms. Figure~\ref{fig:gw_sepctra} shows the amplitude spectrum density (ASD) of the postmerger GWs for the stiff and medium EOS models with different values of $\Gamma_{\rm th}$ and total mass, where the postmerger frequency peaks $f_2$ are highlighted with colored dashed vertical lines. For all the models, the frequency $f_2$ decreases as the thermal index $\Gamma_{\rm{th}}$ increases. This inverse relationship arises primarily due to the decreased compactness of the merger remnant for higher thermal indices. The $f_2$ peak is usually interpreted as the $f$-mode frequency of the remnant MNS, which is broadly proportional to $(M_{\mathrm{MNS}}/R_{\mathrm{MNS}}^3)^{1/2}$, with $M_{\mathrm{MNS}}$ and $R_{\mathrm{MNS}}$ being the typical mass and radius of the remnant MNS. A more compact remnant resulting from lower thermal pressures inherently exhibits higher compactness and, therefore, higher values of $f_2$.

\begin{figure}
    \centering
    \includegraphics[width=0.45\textwidth]{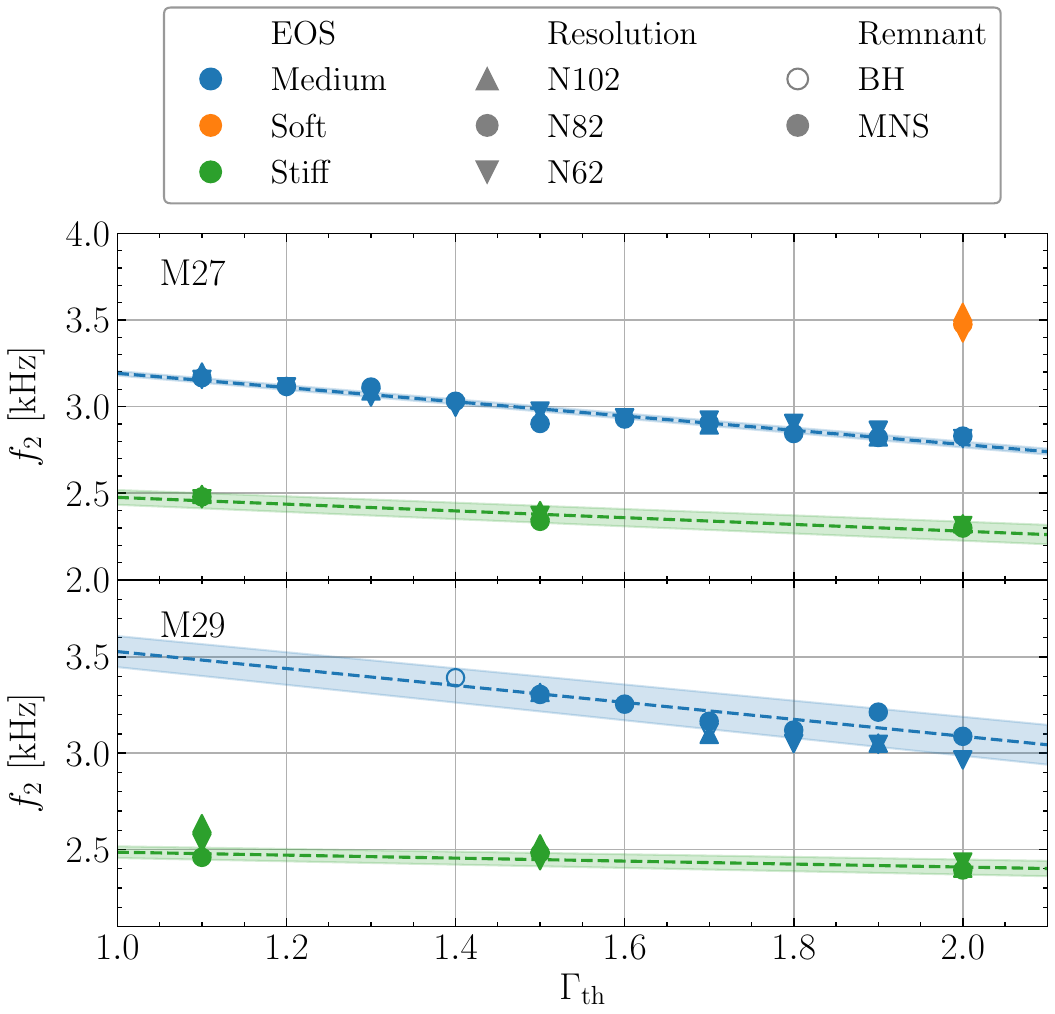}
    \caption{Frequencies of the dominant peak $f_2$ versus $\Gamma_{\rm th}$. The dashed lines represent the best linear fits, while the shaded regions indicate the corresponding standard deviations. The meaning of the legend is the same as Fig.~\ref{fig:rho_max_scatters}.}
    \label{fig:f2_gamma}
\end{figure}

To further quantify the dependence of $f_2$ on the thermal heating efficiency, we plot the relationship between $ f_2$ and $\Gamma_{\rm th}$ in Fig.~\ref{fig:f2_gamma}, excluding the prompt collapse models. For the stiff and medium EOS models, the data are fitted with a linear function of the form $f_2 = a \, \Gamma_{\mathrm{th}} + b$. The corresponding fitting parameters for different EOSs and total masses are summarized in Table~\ref{tab:linear_fits}, with the results also illustrated in Fig.~\ref{fig:f2_gamma}. The slope $a$ characterizes the sensitivity of $f_2$ to the thermal index $\Gamma_{\rm th}$, serving as an indicator of thermal heating efficiency. The absolute value is significantly larger for the medium EOS ($|a|\simeq 0.4$) compared to the stiff EOS ($|a|\lesssim 0.2$), as shock heating has a more pronounced effect in softer EOSs, resulting in a larger shift in $f_2$ for the same change in $\Gamma_{\rm th}$.

Varying the thermal index $\Gamma_{\rm th}$ by 0.5 can lead to a change in $f_2$ of approximately $0.2\,\rm kHz$ for the medium EOS, and about $0.1\,\rm kHz$ and $0.04\,\rm kHz$ for the stiff EOS with total masses of $2.7\,M_{\odot}$ and $2.9\,M_{\odot}$, respectively. If we adopt the empirical relation between $f_2$ and the radius of the premerger NS proposed in Refs.~\cite{Raithel:2022orm,Vretinaris:2019spn}, a change of $f_2$ by $0.2\,\rm kHz$ corresponds to a variation in the NS radius at $1.6\,M_{\odot}$ of approximately $0.3$--$0.4\,\rm km$. This provides an approximate estimate of the uncertainty introduced by the thermal effects. This is obviously non-negligible when compared to the uncertainties in the zero-temperature EOS, thereby making the determination of the EOS from BNS mergers more challenging.

\begin{table}
\centering
\setlength{\tabcolsep}{12pt}
\caption{Linear fitting parameters for $f_2 = a \, \Gamma_{\mathrm{th}} + b$, along with corresponding standard deviations $\sigma_{a}$ and $\sigma_b$.}
\begin{tabular}{lccc}
\toprule
Model & $a\,[\rm kHz]$ & $b\,[\rm kHz]$ & $(\sigma_a,\  \sigma_b)$ \\
\hline
M-M27 & $-0.411$ & $3.603$ & $(0.0019,\ 0.0046)$ \\
M-M29 & $-0.440$ & $3.967$ & $(0.0103,\ 0.0301)$ \\
H-M27 & $-0.195$ & $2.671$ & $(0.0061,\ 0.0152)$ \\
H-M29 & $-0.077$ & $2.563$ & $(0.0043,\ 0.0108)$ \\
\bottomrule
\end{tabular}
\label{tab:linear_fits}
\end{table}

Additionally, we note that the convergence of numerical results across different spatial resolutions is robust, as depicted by the close alignment of $f_2$ values obtained from varying resolutions. 
This consistency reinforces the reliability of our numerical findings and affirms the suitability of our computational approach for accurately capturing the complex interplay between thermal physics and NS EOS characteristics in postmerger GW emission.

\begin{figure}
  \centering
  \includegraphics[width=0.45\textwidth]{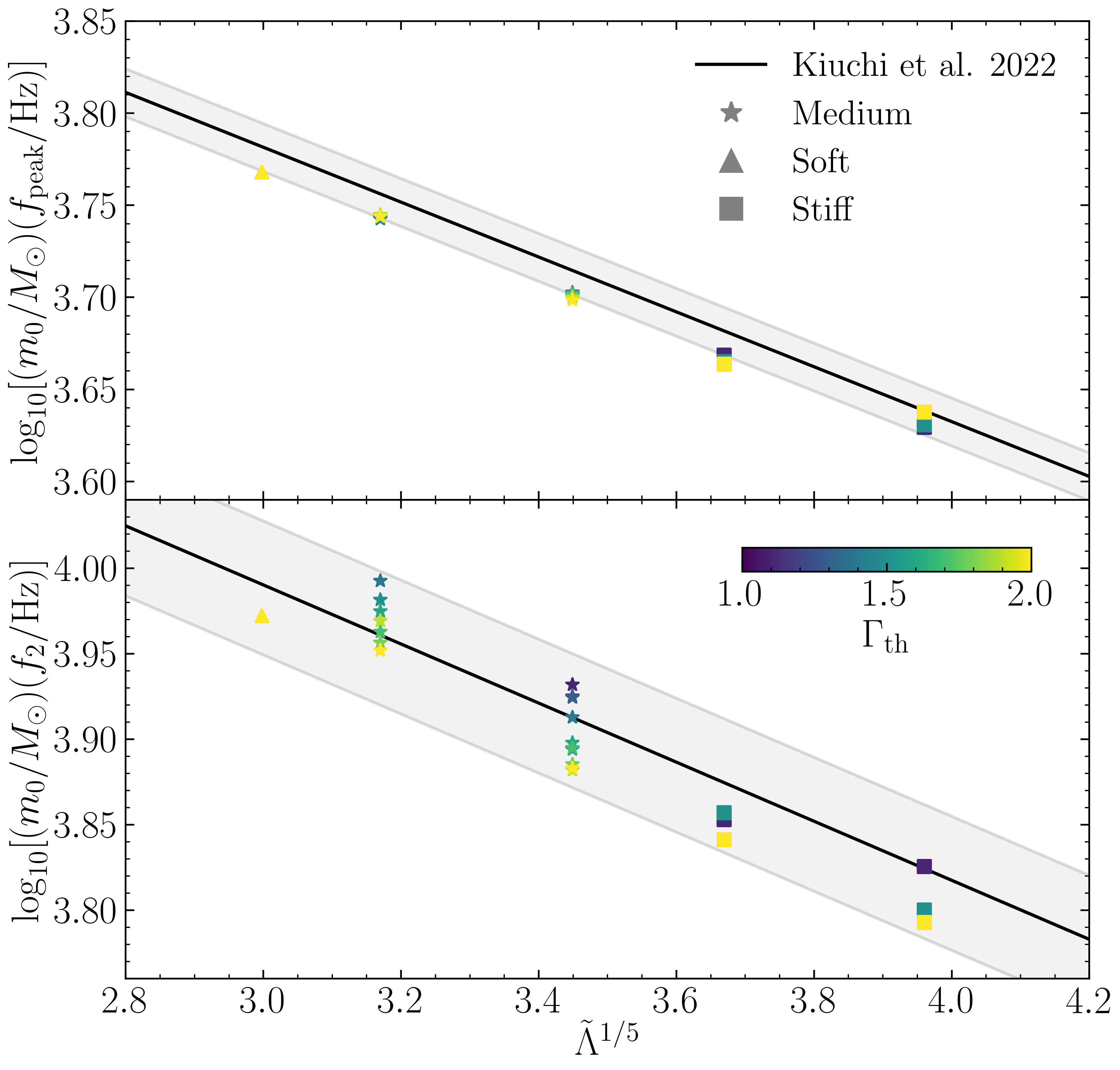}
  \caption{$m_0 f_{\rm peak}$ (upper panel) and $m_0 f_2$ (lower panel) as a function of $\tilde{\Lambda}^{1/5}$, with different markers denoting different EOS and different colors denoting different $\Gamma_{\rm th}$. The black solid lines and the gray shaded regions are the fitting formula and the fitting uncertainty from Ref.~\citep{Kiuchi:2019kzt}.}
  \label{fig:f_univ}
\end{figure}

\begin{figure}
  \centering
  \includegraphics[width=0.45\textwidth]{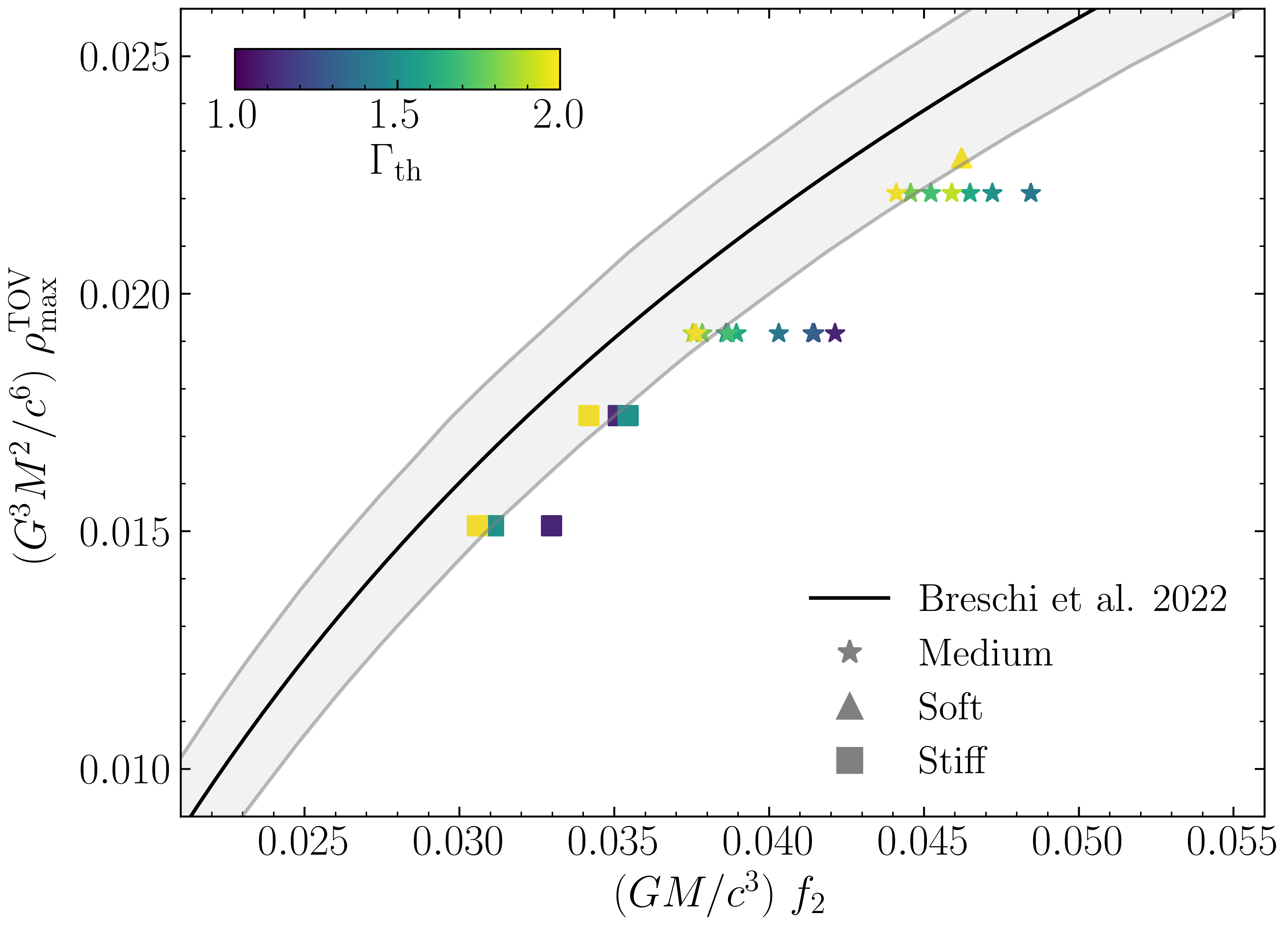}
  \caption{The maximum rest-mass density $\rho^{\rm TOV}_{\rm max}$ of a nonrotating NS at the maximum mass as a function of $f_2$. Different markers/colors denote the different EOSs/$\Gamma_{\rm th}$. The black solid line and the corresponding gray shaded region are the fitting universal relation from Ref.~\citep{Breschi:2021xrx}.}
  \label{fig:f_univ1}
\end{figure}

Quasiuniversal relations linking postmerger GWs to premerger properties provide a useful tool for probing supranuclear matter in NSs~\cite{Bernuzzi:2015rla,Breschi:2019srl,Bauswein:2018bma,Breschi:2021xrx,Blacker:2023afl,Raithel:2022orm,Vretinaris:2019spn}. However, before placing robust constraints, it is crucial to evaluate systematic uncertainties arising from additional underlying physics that can influence the postmerger GW signal. For instance, the impact of uncertain gravity theories has been investigated in Ref.~\cite{Lam:2024azd}. In this work, we focus on assessing the systematics associated with thermal effects.

One class of quasiuniversal relations connects the quantities $m_0 f_{\rm peak}$~\cite{Bernuzzi:2014kca,Bernuzzi:2015rla,Kiuchi:2019kzt} or $m_0 f_2$~\cite{Breschi:2019srl,Kiuchi:2019kzt,Bernuzzi:2015rla} to the mass-weighted dimensionless tidal deformability $\tilde{\Lambda}$, defined as
\begin{equation}
\tilde{\Lambda} = \frac{16}{13} \frac{(m_1 + 12 m_2) m_1^4 \Lambda_1 + (m_2 + 12 m_1) m_2^4 \Lambda_2}{(m_1 + m_2)^5}\,.
\end{equation}
Here, \( m_{1,2} \) and \( \Lambda_{1,2} \) denote the masses and dimensionless tidal deformabilities of the two component NSs, respectively. As shown in Fig.~\ref{fig:f_univ}, we confirm that the variations in these quasiuniversal relations due to changes in $ \Gamma_{\rm th}$ remain within the error bars.  

Reference~\cite{Bauswein:2018bma} showed that the relation between $m_0 f_2$ and $ \tilde{\Lambda}$ can be significantly violated (far outside the error bar) if the BNS merger develops a strong first-order phase transition in the merger remnant. A similar deviation has also been identified in the presence of non-convex dynamics in the merger remnant~\cite{Rivieccio:2024sfm}. Our results suggest that this conclusion is likely robust against thermal effects.
However, it is worth noting that our study suggests the error bars arise not only from uncertainties in the cold EOS, but also from those in the finite-temperature component. This entanglement between the cold and thermal parts complicates the process of constraining the EOS using quasiuniversal relations. 

Reference~\cite{Breschi:2021xrx} identified a quasiuniversal relation between $f_2$ and the central density of the maximum-mass nonrotating NS, providing a new approach to constrain the EOS. As shown in Fig.~\ref{fig:f_univ1}, we find that the violations is much larger than their error bar when $\Gamma_{\rm th}\lesssim 1.5$, with smaller $\Gamma_{\rm th}$ corresponding to larger violations. Accordingly, the conclusion drawn in Ref.~\cite{Breschi:2021xrx}, which suggests that the pressure-density relation up to the maximum mass and the maximum mass of NSs can be inferred from BNS merger signals, may warrant more careful scrutiny.

\section{Summary}\label{sec4}

In this study, we systematically explored the influence of thermal effects in equal-mass BNS mergers using comprehensive numerical relativity simulations. 
To model finite-temperature behavior, we employed a hybrid EOS framework combining a cold EOS, rigorously constrained by multimessenger astrophysical observations and advanced theoretical calculations, with an ideal-gas thermal component described by the thermal index $\Gamma_{\rm th}$. 
By varying $\Gamma_{\rm th}$, we effectively parametrized different efficiencies of the thermal heating. 
Our simulations covered an extensive parameter space, including three distinct EOSs (soft, medium, and stiff) and two binary masses ($m_0 = 2.7~\text{and}~2.9M_\odot$). 
We further validated the numerical accuracy and reliability of our findings through convergence tests performed with different resolutions, i.e., N62, N82, and N102.

Our results revealed that the choice of thermal index significantly affects merger dynamics and the subsequent evolution of the remnant, particularly near critical thresholds delineating stable NS remnants from prompt or delayed BH formation. 
Specifically, systems situated close to the collapse threshold exhibited a strong sensitivity to $\Gamma_{\rm th}$: smaller values of $\Gamma_{\rm th}$ resulted in prompt or delayed collapse, whereas larger ones facilitated the formation of a long-lived MNS.

Regarding dynamical mass ejection, we observed a two-stage ejection process. 
Initially, higher thermal indices lead to greater ejecta mass through enhanced shock heating efficiency. 
In the subsequent phase dominated by torque-driven ejection, smaller thermal indices yielded more compact remnants, thereby amplifying angular momentum transport and boosting ejecta mass. 
However, for a softer EOS combined with small thermal indices, the gravitational trapping effect emerged prominently, suppressing torque-driven ejection efficiency and consequently reducing total ejecta mass. 
This complex interplay resulted in characteristic peaks in ejecta mass as a function of $\Gamma_{\rm th}$, with the location of these peaks shifting depending on the binary mass $m_0$.
Besides, by comparing the ``spurious ejecta mass'' in the inspiral phase and the baryonic mass conservation error, we also found that a piecewise constant $\Gamma_{\rm th}$ hybrid EOS framework with a smaller $\Gamma_{\rm th}$ in the inspiral phase may be more appropriate to reduce the mass dissipation during the inspiral phase.

In the context of GW signals, thermal effects were found to be negligible during the inspiral phase but substantially impacted the postmerger emission. 
Through a detailed spectral analysis of postmerger GWs, we identified a clear inverse relationship between the dominant frequency peak $f_2$ and the thermal index: higher thermal pressures produced less compact remnants, thus lowering the GW frequencies. 
These variations can be non-negligible compared to the uncertainties in the cold EOS, introducing additional challenges for EOS constraints from BNS mergers. For instance, quasi-universal relations connecting post-merger gravitational wave features to premerger properties have been considered promising for EOS inference. However, our results indicate that thermal effects can introduce uncertainties in these relations comparable to those arising from the cold EOS, thereby complicating their application. In particular, the relation between the dominant post-merger frequency $f_2$ and the central density of the maximum-mass nonrotating NS is significantly violated, suggesting that its use for EOS constraints may require further systematic investigation.

While our simulations systematically studied the thermal effects in BNS mergers, several aspects remain to be addressed. 
Future studies should incorporate more sophisticated finite-temperature EOS models derived from microscopic nuclear theory, alongside essential physical processes such as neutrino radiation transport including neutrino heating and cooling, and magnetohydrodynamic effects. 
Incorporating these additional physical mechanisms will further refine theoretical predictions, enhancing our ability to accurately interpret observational data from upcoming multi-messenger astrophysical campaigns.

\acknowledgements

We thank Kota Hayashi, Kyohei Kawaguchi, Hao-Jui Kuan, Takami Kuroda, and Alan Tsz-Lok Lam for useful discussions and valuable comments.
We thank Achim Schwenk for giving us the motivation for this work.
Numerical computations were carried out on Sakura, Raven, and Viper at Max Planck Computing and Data Facility.
This work was in part supported by the National Natural Science Foundation of China under Grants No.~12233011, the Project for Young Scientists in Basic Research (Grant No.~YSBR-088) of the Chinese Academy of Sciences, the General Fund (Grant No.~2023M733735) of the China Postdoctoral Science Foundation (CPSF), and Grant-in-Aid for Scientific Research (Grants No.~23H04900 and No.~23K25869) of Japanese MEXT/JSPS.

\bibliographystyle{apsrev4-2}
\bibliography{bibtex}

\end{document}

%% file: aas_macros.tex
\def\aap{A\&A}                % Astronomy and Astrophysics
\def\nphysa{Nucl.~Phys.~A}   % Nuclear Physics A